\documentstyle[preprint,prd,aps]{revtex}

 \immediate\write16{This is `epsf.tex' v2.7 <25 October 1996>}%
\newread\epsffilein    
\newif\ifepsfatend     
\newif\ifepsfbbfound   
\newif\ifepsfdraft     
\newif\ifepsffileok    
\newif\ifepsfframe     
\newif\ifepsfshow      
\epsfshowtrue          
\newif\ifepsfshowfilename 
\newif\ifepsfverbose   
\newdimen\epsfframemargin 
\newdimen\epsfframethickness 
\newdimen\epsfrsize    
\newdimen\epsftmp      
\newdimen\epsftsize    
\newdimen\epsfxsize    
\newdimen\epsfysize    
\newdimen\pspoints     
\def\epsfbox#1{\global\def\epsfllx{72}\global\def\epsflly{72}%
   \global\def\epsfurx{540}\global\def\epsfury{720}%
   \def\lbracket{[}\def\testit{#1}\ifx\testit\lbracket
   \let\next=\epsfgetlitbb\else\let\next=\epsfnormal\fi\next{#1}}%
\def\epsfgetlitbb#1#2 #3 #4 #5]#6{%
   \epsfgrab #2 #3 #4 #5 .\\%
   \epsfsetsize
   \epsfstatus{#6}%
   \epsfsetgraph{#6}%
}%
\def\epsfnormal#1{%
    \epsfgetbb{#1}%
    \epsfsetgraph{#1}%
}%
\def\epsfgetbb#1{%
    \openin\epsffilein=#1
    \ifeof\epsffilein
        \errmessage{Could not open file #1, ignoring it}%
    \else                       
        {
            \chardef\other=12
            \def\do##1{\catcode`##1=\other}%
            \dospecials
            \catcode`\ =10
            \epsffileoktrue         
            \epsfatendfalse     
            \loop               
                \read\epsffilein to \epsffileline
                \ifeof\epsffilein 
                \epsffileokfalse 
            \else                
                \expandafter\epsfaux\epsffileline:. \\%
            \fi
            \ifepsffileok
            \repeat
            \ifepsfbbfound
            \else
                \ifepsfverbose
                    \immediate\write16{No BoundingBox comment found in %
                                    file #1; using defaults}%
                \fi
            \fi
        }
        \closein\epsffilein
    \fi                         
    \epsfsetsize                
    \epsfstatus{#1}%
}%
\def\epsfclipon{\def\epsfclipstring{ clip}}%
\def\epsfclipoff{\def\epsfclipstring{\ifepsfdraft\space clip\fi}}%
\epsfclipoff 
\def\epsfspecial#1{%
     \epsftmp=10\epsfxsize
     \divide\epsftmp\pspoints
     \ifnum\epsfrsize=0\relax
       \includegraphics{\ifepsfdraft}%
     \else
       \epsfrsize=10\epsfysize
       \divide\epsfrsize\pspoints
       \includegraphics{\ifepsfdraft}%
     \fi
}%
\def\epsfframe#1%
{%
  \leavevmode                   
  \setbox0 = \hbox{#1}%
  \dimen0 = \wd0                                
  \advance \dimen0 by 2\epsfframemargin         
  \advance \dimen0 by 2\epsfframethickness      
  \vbox
  {%
    \hrule height \epsfframethickness depth 0pt
    \hbox to \dimen0
    {%
      \hss
      \vrule width \epsfframethickness
      \kern \epsfframemargin
      \vbox {\kern \epsfframemargin \box0 \kern \epsfframemargin }%
      \kern \epsfframemargin
      \vrule width \epsfframethickness
      \hss
    }
    \hrule height 0pt depth \epsfframethickness
  }
}%
\def\epsfsetgraph#1%
{%
   \leavevmode
   \hbox{
     \ifepsfframe\expandafter\epsfframe\fi
     {\vbox to\epsfysize
     {%
        \ifepsfshow
            \vfil
            \hbox to \epsfxsize{\epsfspecial{#1}\hfil}%
        \else
            \vfil
            \hbox to\epsfxsize{%
               \hss
               \ifepsfshowfilename
               {%
                  \epsfframemargin=3pt 
                  \epsfframe{{\tt #1}}%
               }%
               \fi
               \hss
            }%
            \vfil
        \fi
     }%
   }}%
   \global\epsfxsize=0pt
   \global\epsfysize=0pt
}%
\def\epsfsetsize
{%
   \epsfrsize=\epsfury\pspoints
   \advance\epsfrsize by-\epsflly\pspoints
   \epsftsize=\epsfurx\pspoints
   \advance\epsftsize by-\epsfllx\pspoints
   \epsfxsize=\epsfsize{\epsftsize}{\epsfrsize}%
   \ifnum \epsfxsize=0
      \ifnum \epsfysize=0
        \epsfxsize=\epsftsize
        \epsfysize=\epsfrsize
        \epsfrsize=0pt
      \else
        \epsftmp=\epsftsize \divide\epsftmp\epsfrsize
        \epsfxsize=\epsfysize \multiply\epsfxsize\epsftmp
        \multiply\epsftmp\epsfrsize \advance\epsftsize-\epsftmp
        \epsftmp=\epsfysize
        \loop \advance\epsftsize\epsftsize \divide\epsftmp 2
        \ifnum \epsftmp>0
           \ifnum \epsftsize<\epsfrsize
           \else
              \advance\epsftsize-\epsfrsize \advance\epsfxsize\epsftmp
           \fi
        \repeat
        \epsfrsize=0pt
      \fi
   \else
     \ifnum \epsfysize=0
       \epsftmp=\epsfrsize \divide\epsftmp\epsftsize
       \epsfysize=\epsfxsize \multiply\epsfysize\epsftmp
       \multiply\epsftmp\epsftsize \advance\epsfrsize-\epsftmp
       \epsftmp=\epsfxsize
       \loop \advance\epsfrsize\epsfrsize \divide\epsftmp 2
       \ifnum \epsftmp>0
          \ifnum \epsfrsize<\epsftsize
          \else
             \advance\epsfrsize-\epsftsize \advance\epsfysize\epsftmp
          \fi
       \repeat
       \epsfrsize=0pt
     \else
       \epsfrsize=\epsfysize
     \fi
   \fi
}%
\def\epsfstatus#1{
   \ifepsfverbose
     \immediate\write16{#1: BoundingBox:
                  llx = \epsfllx\space lly = \epsflly\space
                  urx = \epsfurx\space ury = \epsfury\space}%
     \immediate\write16{#1: scaled width = \the\epsfxsize\space
                  scaled height = \the\epsfysize}%
   \fi
}%
{\catcode`\%=12 \global\let\epsfpercent=
\global\def\epsfatend{(atend)}%
\long\def\epsfaux#1#2:#3\\%
{%
   \def\testit{#2}
   \ifx#1\epsfpercent           
       \ifx\testit\epsfbblit    
            \epsfgrab #3 . . . \\%
            \ifx\epsfllx\epsfatend 
                \global\epsfatendtrue
            \else               
                \ifepsfatend    
                \else           
                    \epsffileokfalse
                \fi
                \global\epsfbbfoundtrue
            \fi
       \fi
   \fi
}%
\def\epsfempty{}%
\def\epsfgrab #1 #2 #3 #4 #5\\{%
   \global\def\epsfllx{#1}\ifx\epsfllx\epsfempty
      \epsfgrab #2 #3 #4 #5 .\\\else
   \global\def\epsflly{#2}%
   \global\def\epsfurx{#3}\global\def\epsfury{#4}\fi
}%
\def\epsfsize#1#2{\epsfxsize}%
\tightenlines
\epsfclipon          

\newcommand{\be}{\begin{equation}}
\newcommand{\ee}{\end{equation}}
\def\alt{\ {\raise-3pt\hbox{$\sim$}}\!\!\!\!\!{\raise2pt\hbox{$<$}}\ }
\def\agt{\ {\raise-3pt\hbox{$\sim$}}\!\!\!\!\!{\raise2pt\hbox{$>$}}\ }
\newcommand{\met}{\hbox{{$E_T$}\kern-1.1em\hbox{/}}\kern+0.55em}

\begin{document}

\preprint{\vbox{
\hbox{CTP-TAMU-38-98}
\hbox{OSU-HEP-98-7}
}}

\title{The signature at the Tevatron for the light doubly charged Higgsino
of the supersymmetric left-right model}
\author{B. Dutta$^1$, R. N. Mohapatra $^2$ and D. J. Muller$^3$ }
\address{$^1$ Department of Physics, Texas A\&M University, 
College Station, TX 77840}
\address{$^2$ Department of Physics, University of Maryland, College Park,
MD-20742}
\address{$^3$ Department of Physics, Oklahoma State University, Stillwater, OK
74078}
\date{October, 1998}
\maketitle

\begin{abstract} 
We analyze the signal at the Tevatron
for the doubly charged Higgs superfield of supersymmetric left-right models
with gauge mediated supersymmetry breaking. Due to the
presence of a new coupling, the lighter stau is usually less massive than
the lightest neutralino and is frequently the NLSP\@.
The fermionic
part of the doubly charged Higgs field decays dominantly into two
$\tau$ leptons plus missing energy (gravitino).  We find that the inclusive
two $\tau$-jets  and three $\tau$-jets plus
missing energy signatures could be observable at Run II of the Tevatron.
We also find
that the distribution in angle between the two highest $E_T$ $\tau$-jets
when they come from same sign $\tau$ leptons can be used to
distinguish this model from the others. 
\end{abstract} 


\section{Introduction}

Due to its many attractive features, supersymmetry (SUSY) has become a
primary focus of experimental investigations. Most of the experimental
searches for SUSY so far have been performed in the context of the Minimal
Supersymmetric Standard Model (MSSM)\cite{gunion}. It is important, 
however, to
determine what the signatures are for SUSY models beyond the MSSM\@.
In this work we consider SUSY left-right (LR)
models with gauge mediated supersymmetry breaking (GMSB).

Supersymmetric left-right models (SUSYLR) where the $SU(2)_R$ gauge
symmetry is broken by triplet Higgs fields $\Delta^c$ with $B-L=2$ have many
attractive features:
1) they imply automatic conservation of baryon and lepton number \cite{moh};
2) they provide a natural  solution to the strong and weak CP problems of
the MSSM \cite{rasin}; 3) they yield a natural embedding of the see-saw
mechanism for small  neutrino masses \cite{gell} where the right-handed
triplet  field ($\Delta^c$) that breaks the $SU(2)_R$ symmetry also gives
heavy mass to the right-handed
Majorana neutrino needed for implementing the see-saw mechanism.
 
Recently it has been shown that the doubly charged components of the
 triplet Higgs fields are massless
unless there are some  higher dimensional operators (HDO)
\cite{kuchi,aulakh,chacko,goran2}.  This is independent  of how 
supersymmetry breaking is transmitted to the visible sector 
({\it i.e.},  whether 
it is gravity mediated or it is gauge mediated) and also of whether the 
hidden sector supersymmetry breaking scale is above or  below the
$W_R$ scale. In the presence of HDO's, they  acquire masses of order 
$\sim v^2_R/M_{\rm Pl}$.  Since  the measurement of the  Z-width at LEP 
and SLC
implies that such particles  must  have a mass of at least 45\,GeV, this
puts a lower limit on the $W_R$ scale of about
$10^{10}$\,GeV or so. 
For $W_R$ near this lower limit, the masses of the doubly charged
particles are in the 100\,GeV range.
The rest of the particle spectrum
below the $W_R$ scale can be the same as that of the
MSSM with a massive neutrino or it can have an extra pair of Higgs doublets 
in the 10\,TeV  range depending on the structure of the model.

The ordering of the sparticles masses in the SUSYLR model makes the SUSY
signature distinctive from other SUSY models. Of primary importance to the
signal is the identity of the sparticle which is the next to lightest
SUSY particle (NLSP). Here the NLSP can be the lightest neutralino, the
lighter stau, or the light doubly charged Higgsino (which we henceforth 
call the deltino).
Mass spectra for this type of model have been studied previously \cite{dm}.
It was found that one of the Higgs fields has a coupling with the third
generation charged leptons which reduces the third generation charged
slepton masses. Because of this, in SUSYLR models with GMSB the lighter
stau is predominantly the NLSP whenever the deltino 
is too massive to play that role. As a result, the decay chains of the SUSY
particles typically lead to the lighter stau. The $\tilde{\tau}_1$ then
decays into a $\tau$ lepton and a gravitino ($\tilde{G}$)
which escapes the detector
undetected (leading to missing energy). Since the gravitino mass is on the
order of eV, the emitted $\tau$ will have high $p_T$ enhancing its
detection possibility. Moreover, pair production of the light doubly charged
Higgsinos always produces four $\tau$ leptons. When the $\tilde{\tau}_1$ is
the NLSP, this occurs through
$\tilde{\Delta}^{c\pm \pm} \rightarrow \tilde{\tau}_1^\pm \tau^\pm$
followed by $\tilde{\tau}_1 \rightarrow \tau \, \tilde{G}$.
When the $\tilde{\Delta}^{c\pm \pm}$ is the NLSP, this occurs through
the stau mediated decay
$\tilde{\Delta}^{c\pm \pm} \rightarrow \tau^\pm \tau^\pm \tilde{G}$.
One can get 
a similar
signal in supergravity motivated LR models with
the gravitinos replaced by the lightest neutralino
(its mass is greater than 33\,GeV),
which will constitute the
missing energy.

Signals involving two or more high $p_T$ $\tau$ leptons are also 
important signals for conventional GMSB models as the lighter stau is 
frequently the NLSP for these models as well. In the SUSYLR model, however,
we find that the production of the deltino can greatly
enhance the signal. In addition, since the deltino decays
into like sign $\tau$ leptons, we find that the distribution in angle
between same sign $\tau$ leptons can be used to distinguish this model
from other GMSB models.

In this paper we analyze in detail the signal for SUSY production in the
context of the SUSYLR model. First, we discuss the sparticle masses and
the production cross sections for the major SUSY production modes:
$\chi_1^+ \chi_1^-$ production, $\chi_1^\pm \, \chi_2^0$ production,
slepton pair production and $\tilde{\Delta}^{c++} \tilde{\Delta}^{c--}$
production. Next, we analyze the $\tau$-jet signal. We give the typical
two, three and four $\tau$-jet inclusive production cross sections with
and without standard cuts. Last, we show how the 
distribution in angle between
the $\tau$-jets based on their charges can distinguish this model from other
models whose signals also involve the production of $\tau$ leptons.

\section{Sparticle Masses and Production}

Since the observed signal depends on the masses of the sparticles, we first
begin by describing the model and the corresponding mass spectrum.
The particle content of this model above the LR scale includes $\phi(2,2,0)$,
$\Delta(3,1,2)$, $\bar \Delta(3,1,-2)$, $\Delta^c(1,3,-2)$, 
$\bar \Delta^c(1,3,-2)$ and a singlet where the numbers in the parentheses 
refer to their transformation properties under
$SU(2)_L\times SU(2)_R\times U(1)_{B-L}$.
The LR symmetric  superpotential for this theory is
\begin{eqnarray}  W & = & {\bf h}^{(i)}_q Q^T \tau_2 \Phi_i \tau_2 Q^c + {\bf
h}^{(i)}_l L^T \tau_2 \Phi_i \tau_2 L^c
\nonumber\\
  & +  & i ( {\bf f} L^T \tau_2 \Delta L + {\bf f}_c {L^c}^T \tau_2 \Delta^c 
L^c)
\nonumber\\
  & +  & M_{D} [{\rm Tr} ( \Delta \bar{\Delta} ) +
 {\rm Tr} ( \Delta^c \bar{\Delta}^c )] +\lambda S(\Delta\overline{\Delta}
-\Delta^c\overline{\Delta^c}) + \mu_S S^2 \nonumber\\
 & + & 
\mu_{ij} {\rm Tr} ( \tau_2 \Phi^T_i \tau_2 \Phi_j )+ W_{\it NR}
\label{eq:superpot}
\end{eqnarray}
where $W_{\it NR}$ denotes nonrenormalizable terms arising from
higher scale physics such as grand unified theories or Planck scale effects.
Demanding that the $F_{\Delta}$, $F_s$ and $F_{\Delta^c}$ terms vanish,
it has been found \cite {dm} that the VEV of S becomes $M_{D}\over \lambda$.
The VEVs for the $\Delta$ and $\bar\Delta$ vanish and their masses become of the
order of $2 M_{D}$.  
Let us now see what happens to the different components of the 
$\Delta^c$ fields. One linear combination of the singly
charged and neutral  Higgs fields pick up a mass on the order of $v_R$. The
doubly charged Higgsinos are massless unless the nonrenormalizable terms
are introduced. In the presence of $W_{NR}$,
the doubly charged fields acquire masses of order $v_R^2/M_{\rm Pl}$.
This result is true when the vacuum conserves R-parity. When the vacuum 
breaks R-parity, the scale of the right handed $W$'s has an upper bound on
the order of a few TeV\@. Therefore, in this version of the model, all 
particles, including the doubly charged bosons and fermions, have masses
also in the few hundred GeV to TeV range. In what follows we work with the
R-parity conserving version for simplicity.

After integrating out the fields at the left-right scale, we are left with
the following additional part to the MSSM:
\begin{equation}
W=M_{\Delta}\Delta^{c--}\bar{\Delta}^{c++}+f_il^cl^c\Delta^{c--}
\end{equation}  
where we have assumed that f is diagonal. The PSI esperiment \cite{psi} 
has put an upper bound on the product of the first two generation
couplings of
$f_1f_2<1.2\times 10^{-3}$. The magnitude of $f_3$ is unrestrained. The
term $M_{\Delta}$ originates from the nonrenormalizable terms.

The model considered here involves gauge mediated supersymmetry breaking.
In GMSB type models the SUSY breaking is communicated to the observable
sector by the SM gauge interactions \cite{dine}. We choose GMSB since
the lighter stau will then have a mass that is almost always below that
of the lightest neutralino. The lighter stau then decays to a $\tau$ lepton
and a gravitino. Since the gravitino  is  very light,
the $\tau$ lepton will typically be very energetic.

In the 
GMSB model, the sparticle spectrum depends on the following parameters:
$M$, $\Lambda$, $n$, $\tan \beta$, $f_3$, $M_{\tilde{\Delta}} (M)$ and 
the sign of $\mu$. 
$M$ is the messenger scale. The parameter $n$ is dictated by the choice of 
the vector-like messenger sector. In this calculation we will assume 
that each flavor in
the messenger sector consists of a vector like isosinglet pair of 
fields ($Q+\bar Q$) and a vector like weak isodoublet pair $L+\bar L$.   The
definition of
$\tan \beta$ is that $\tan \beta \equiv v_2/v_1$ where $v_2$ is the VEV
for the up-type ($H_u$) Higgs doublet and $v_1$ is the VEV for the
down-type ($H_d$) Higgs doublet. 
$M_{\tilde{\Delta}} (M)$ is the messenger scale value for the
deltino mass.
The parameter $\mu$ is the coefficient
in the bilinear mixing term, $\mu H_u H_d$, in the superpotential.
Constraints coming from $b \rightarrow s \gamma$ strongly favor negative
values for $\mu$ \cite{dwt} and, in the cases considered in this work, 
$\mu$ is taken to be negative. Demanding that the EW symmetry be broken 
radiatively fixes the magnitude of $\mu$ and the parameter $B$ (from the
$B \mu H_u H_d$ term in the scalar potential) in terms of the other
parameters of the theory.

The soft SUSY breaking gaugino and scalar masses at the messenger scale
are given by \cite{martin}
\be
\label{gmass}
\tilde{M}_i (M) = n g \left ( \frac{\Lambda}{M} \right )
\frac{\alpha_i(M)}{4 \pi} \Lambda
\ee
and
\be
\label{smass}
\tilde{m}^2 (M) = 2 n f \left ( \frac{\Lambda}{M} \right )
    \sum_{i = 1}^{3} k_i C_i \left ( \frac{\alpha_i(M)}{4 \pi} \right )^2
    \Lambda^2
\ee
where the $\alpha_i$ are the three SM gauge couplings and
$k_i =$ 1, 1 and 3/5 for SU(3), SU(2) and U(1), respectively.
The $C_i$ are zero for gauge singlets and are 4/3, 3/4 and ($Y$/2)$^2$
for the fundamental representations of SU(3), SU(2) and U(1), respectively
(with $Y$ given by $Q = I_3 + Y/2$). $g(x)$ and $f(x)$ are messenger scale
threshold functions.

We calculate the SUSY mass spectrum by using the appropriate renormalization
group equations \cite{barger}. We first run the Yukawa couplings (including 
the three new couplings $f_{1,2,3}$) and the gauge couplings from the
weak scale up to the
messenger scale. At the messenger scale, we apply the boundary conditions
given by the equations above and then use the RGEs for the soft SUSY
couplings and masses in order to run down to the weak scale.

The mass spectrum here is much like that expected in minimal GMSB models.
The gravitino is always the LSP\@. Since SUSY breaking is communicated to
the visible sector by gauge interactions, the mass differences between the
superparticles depend on their gauge interactions. This creates a
hierarchy in mass between electroweak and strongly interacting sparticles.
Eq.~\ref{gmass} shows that the gluino is more massive than the charginos
and neutralinos, while Eq.~\ref{smass} shows that the squarks are
considerably more massive than the sleptons.
Thus in minimal GMSB models,
the lightest neutralino and the lighter stau fight for the NLSP spot
\cite{dwt,gmsb}.
In this model, the deltino also joins the race to become the NLSP\@.

We will concentrate the analysis on those regions of the parameter space
where either the lighter stau or the deltino is the NLSP\@.
Whether or not the deltino is the NLSP depends on the mass it gets from
the higher dimensional terms. If this mass is too high, then either the
$\tilde{\tau}_1$ or $\chi_1^0$ is the NLSP\@. 
The lighter stau can be much lighter in our SUSYLR model than in
conventional GMSB models due to the presence of the additional 
coupling $f_3$.
Thus the lighter stau will be lighter than the $\chi_1^0$ for a larger region
of the parameter space and the $\tilde{\tau}_1$ has a greater
potential to be the NLSP in this SUSYLR model.

There are a number of potential SUSY production mechanisms here. Given the
current lower bounds on the various sparticle masses and the hierarchy
of sparticle masses in GMSB models, the important SUSY production
mechanisms will typically include EW gaugino production. At the Tevatron,
chargino pair ($\chi_1^+ \chi_1^-$) production takes place through
s-channel $Z$ and $\gamma$ exchange and $\chi_2^0 \, \chi_1^\pm$
production is through s-channel $W$ exchange. Squark exchange via the
t-channel also contributes to both processes, but the contributions are
expected to be negligible since the squark masses are large in GMSB
models. The production of $\chi_1^0 \, \chi_1^\pm$ is suppressed due to the
smallness of the coupling involved.

In addition to these usual SUSY production mechanisms of the MSSM, we also 
have deltino
pair ($\tilde{\Delta}^{c++} \tilde{\Delta}^{c--}$) production. This proceeds
through s-channel $Z$ and $\gamma$ exchange. Given that the
$\tilde{\Delta}^{c\pm \pm}$ can be relatively light, it can be a very
important SUSY production mode. In fact, it frequently is the dominant 
mode.

The possible final state configurations at the Tevatron depend on the
sparticle spectrum and on which SUSY production mode is dominant, but they
will have certain aspects in common. When the $\tilde{\tau}_1$ is the NLSP,
the various possible decay modes will (usually) produce at least two
$\tau$  leptons arising from the decays of the lighter staus. In addition,
there can also be large \met\ due to the stable gravitinos and neutrinos
escaping detection. When the deltino is the NLSP, the standard
SUSY production modes of EW gauginos can still produce large
numbers of $\tau$-jets if the $\tilde{\tau}_1$ is the next to next 
to lightest SUSY particle (in which case the $\tilde{\tau}_1$ is lighter 
than the $\chi_1^0$) so that the 
decay chains of the sparticles will still lead to the $\tilde{\tau}_1$.

Pair production of the deltino leads to copious quantities of $\tau$ leptons
irrespective of what the NLSP is. This is because the 
deltino couples to
leptons/sleptons but not to quarks/squarks. In addition, the coupling to
the third generation can be much greater than the small coupling to the
1st and 2nd generations. 
Thus, when the $\tilde{\tau}_1$ is the NLSP, the deltino decays via
$\tilde{\Delta}^{c\pm \pm} \rightarrow \tilde{\tau}^\pm_1 \tau^\pm$ with 
the stau decaying via $\tilde{\tau}_1 \rightarrow \tau \, \tilde{G}$.
On the other hand, when the deltino is the NLSP, it decays via 
the $\tilde{\tau}$ mediated three-body decay mode
$\tilde{\Delta}^{c\pm \pm} \rightarrow \tau^\pm \tau^\pm \tilde{G}$. Thus
$\tilde{\Delta}^{c\pm \pm}$ pair production generally results in the
production of four $\tau$ leptons (two from each deltino).

\section{Tau Jet Analysis}

As mentioned above, SUSY production for this SUSYLR model leads to the
production of copious quantities of $\tau$ leptons. $\tau$ leptons
are typically identified at colliders by their hadronic decays to
thin jets. 
We now give a detailed account of the possible $\tau$-jet signatures
for SUSY production at the Tevatron in the context of the left-right
GMSB model.

This analysis is performed in the context of the Main Injector (MI) and
TeV33 upgrades of the Tevatron collider. The center of mass energy is
taken to be $\sqrt{s} = 2$\,TeV and the integrated luminosity is taken to
be 2\,fb$^{-1}$ for the MI upgrade and 30\,fb$^{-1}$ for the TeV33
upgrade.

In performing this analysis, the cuts employed are that final state charged
leptons must have $p_T > 10$\,GeV and a pseudorapidity,
$\eta \equiv -\ln ( \tan\frac{\theta}{2} )$ where $\theta$ is the polar
angle with respect to the proton beam direction, of magnitude less than 1.
Jets must have $E_T > 10$\,GeV and $|\eta| < 2$. In addition, hadronic
final states within a cone size of
$\Delta R \equiv \sqrt{ (\Delta \phi)^2 + (\Delta \eta)^2 } = 0.4$ are
merged to a single jet. Leptons within this cone radius of a jet are
discounted. For a $\tau$-jet to be counted as such, it must have
$|\eta| < 1$. The most energetic $\tau$-jet is required to have
$E_T > 20$\,GeV\@. In addition, a missing transverse energy cut of
\met\ $> 30$\,GeV is imposed.

\begin{figure}
\centering
\epsfxsize=0.98\textwidth
\epsfbox[68 124 494 545]{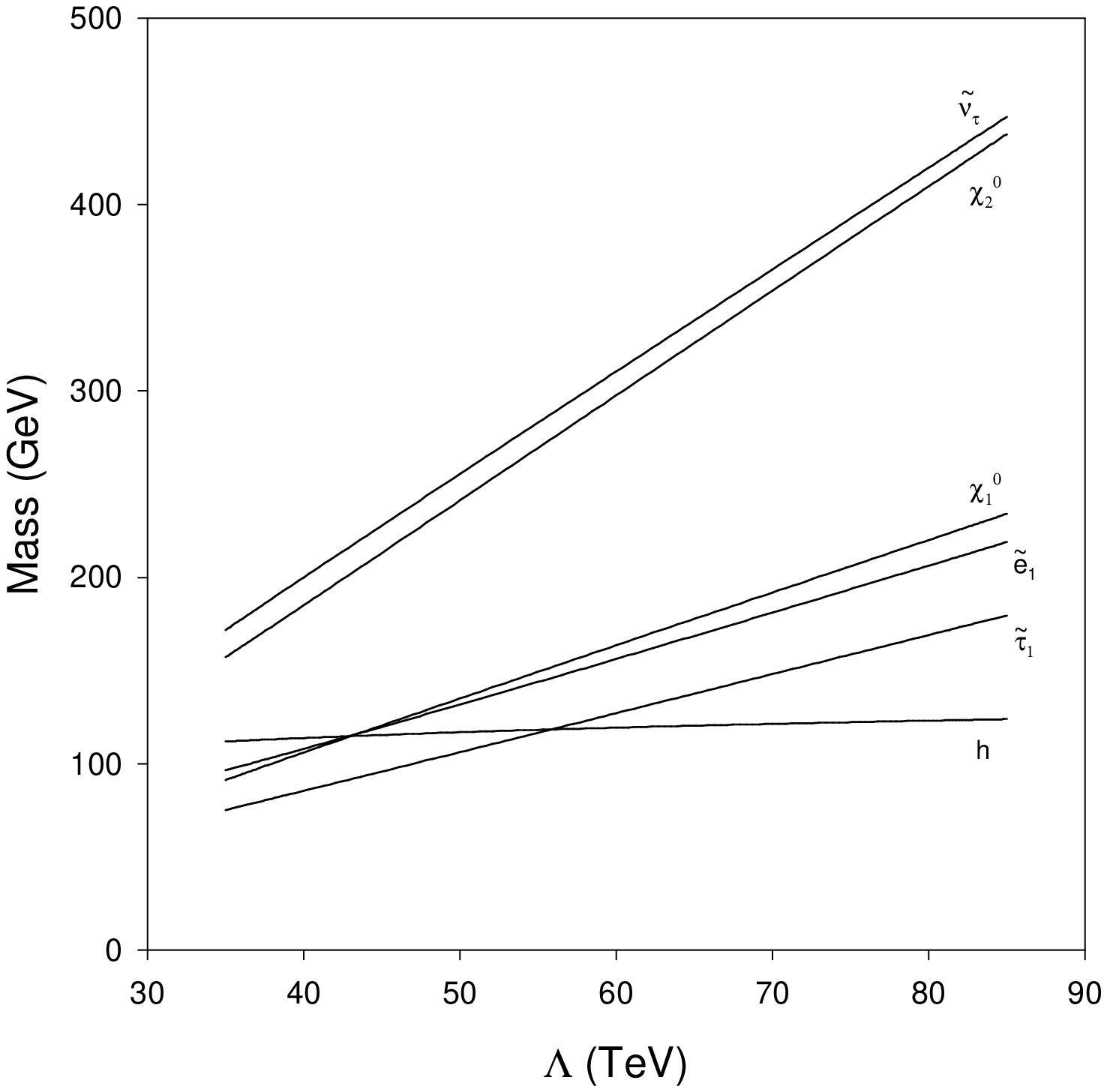}
\vskip 0.35cm
\caption{\label{mass} Masses of the particles of interest for the input
parameters $\tan \beta = 15$, $M$/$\Lambda$ = 3, $n = 2$,
$f_3 = 0.5$, $f_2 = 0.05$, $f_1 = 0.05$ and 
$M_{\tilde{\Delta}} (M) = 90$\,GeV.}
\end{figure}

\begin{figure}
\centering
\epsfxsize=0.98\textwidth
\epsfbox[68 125 494 546]{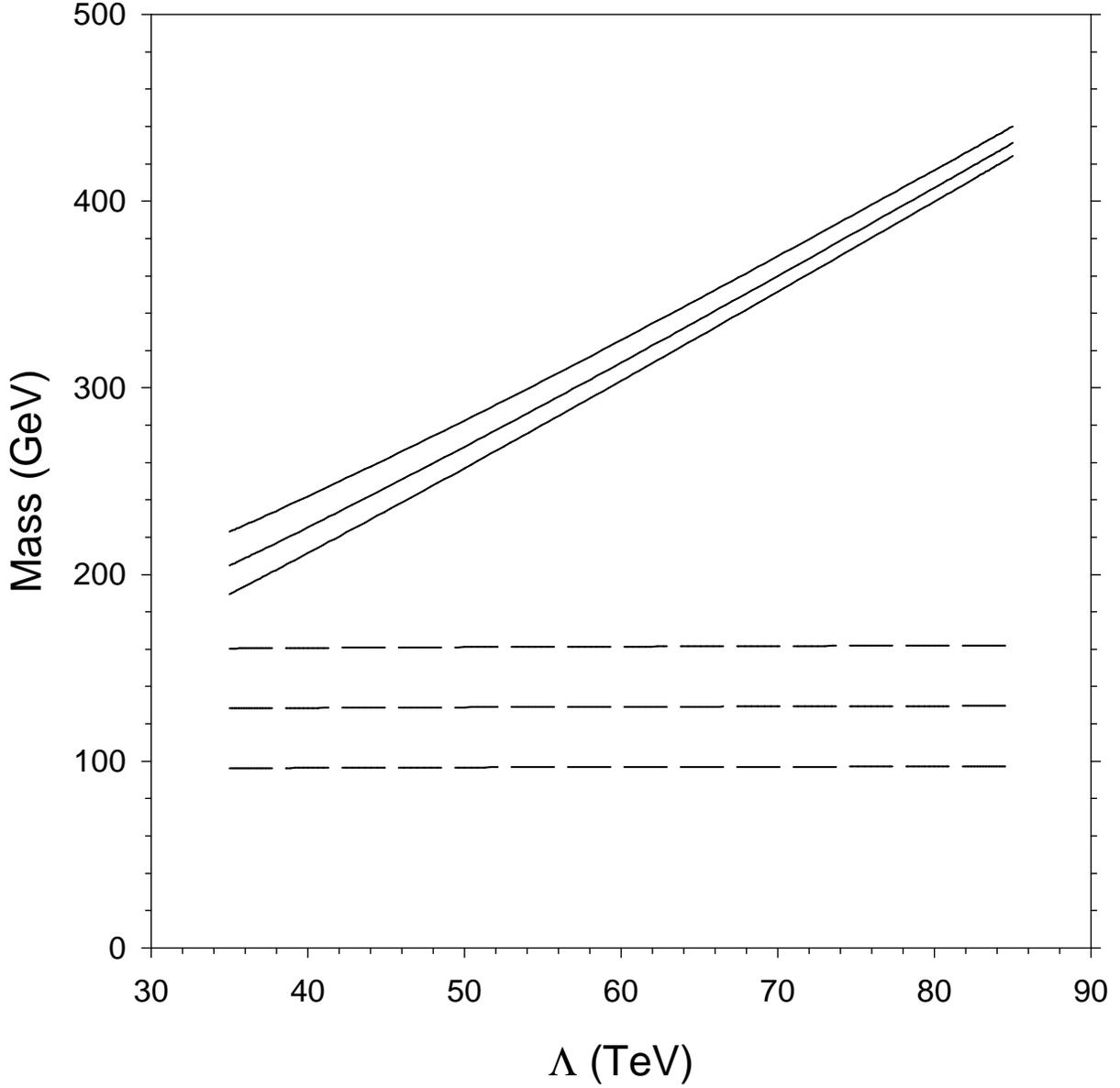}
\vskip 0.35cm
\caption{\label{delmass} Masses of the delta and deltino. The dashed lines
represent the deltino, while the solid lines represent the delta boson.
The parameters used are $\tan \beta = 15$, $n = 2$ and $M$/$\Lambda$ = 3.
From bottom to top, the lines in each set are for a messenger scale deltino
mass of 90, 120 and 150\,GeV\@.}
\end{figure}

The signatures for SUSY production depend on the hierarchy of sparticle
masses. This, in turn, depends on the values the parameters of
the theory takes. The parameters considered in this analysis are
$\tan \beta = 15$, $n = 2$, $M$/$\Lambda$ = 3, $f_3 = 0.5$, $f_2 = 0.05$
and $f_1 = 0.05$. We vary $\Lambda$ from 35 to 85\,TeV\@.
For the messenger scale deltino mass, we use the values 90, 120 and
150\,GeV\@. The masses of some of the particles of interest are given
in Fig.~\ref{mass} and Fig.~\ref{delmass}.  In Fig.~\ref{mass} we take
$M_{\tilde{\Delta}} (M) = 90$\,GeV, but the masses of the gauginos and
sleptons (with the exception of the stau) do not vary much with the
messenger scale deltino mass. Fig.~\ref{delmass} gives the masses of the
delta boson and the deltino. The deltino mass is not very sensitive to
the value of $\Lambda$, while the delta boson mass is highly dependent
on $\Lambda$ due to the contributions from the messenger scale loops 
(which give it mass along with the nonrenormalizable part).
Given the substantially higher $\Delta^c$ boson mass, $\Delta^c$ production
is not very important at the Tevatron.

\begin{figure}
\centering
\epsfxsize=0.98\textwidth
\epsfbox[68 124 494 545]{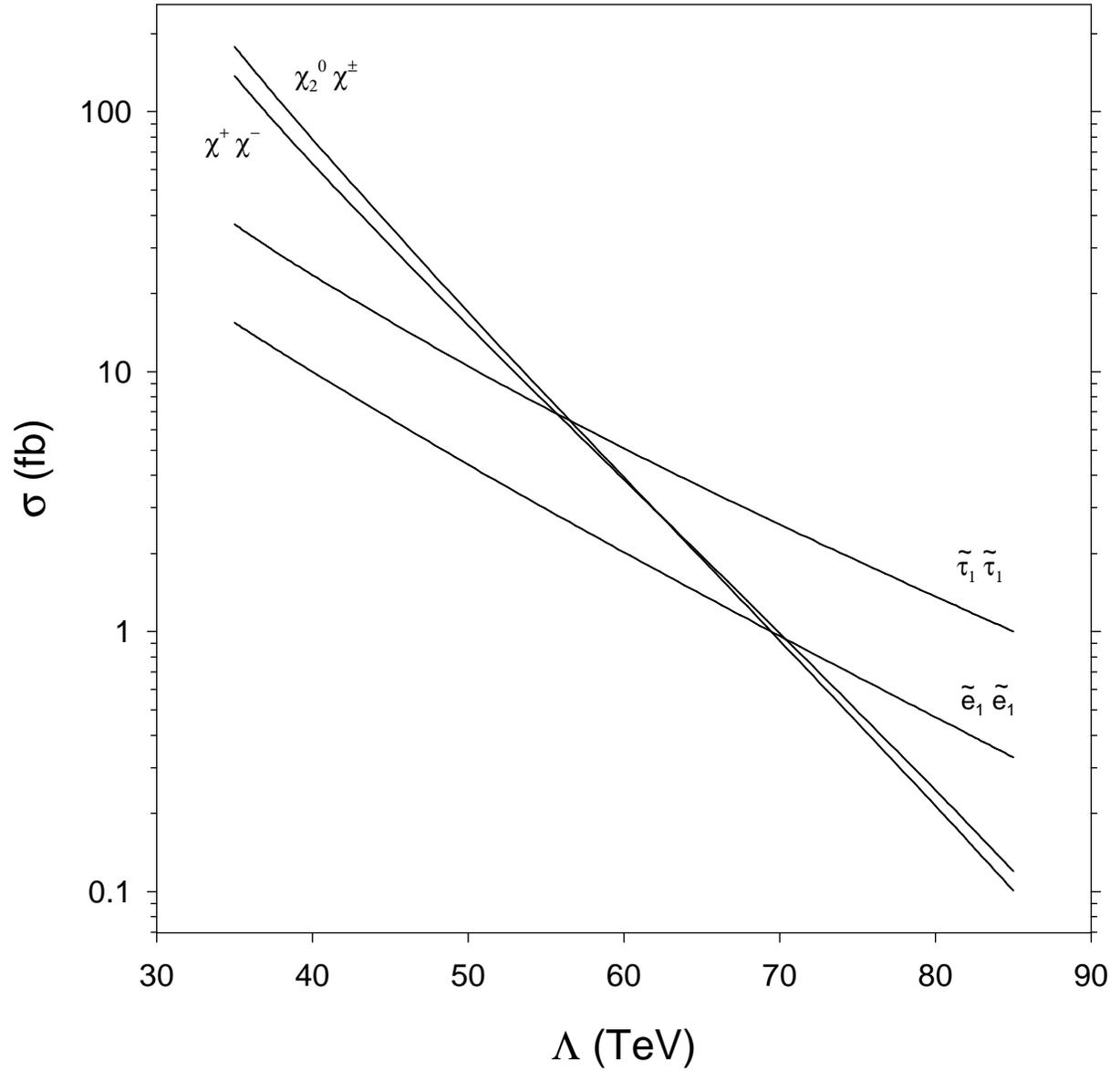}
\vskip 0.35cm
\caption{\label{cross} Cross sections for the standard SUSY production modes
for the parameters $\tan \beta = 15$, $M$/$\Lambda$ = 3,
$n = 2$, $f_3 = 0.5$, $f_2 = 0.05$, $f_1 = 0.05$ and 
$M_{\tilde{\Delta}} (M) = 90$\,GeV.}
\end{figure}

There are several potential SUSY production modes here.
The cross sections for the more traditional SUSY production modes are given
in Fig.~\ref{cross}. 
We also have deltino pair production; the cross sections for which are
tabulated in
Table~\ref{delcross}. Since the deltino mass does not vary much over the
values of $\Lambda$ considered, the cross section for deltino pair
production does not vary much either. This cross section is high enough
for all the deltino masses considered that deltino pair production is
always an important SUSY production mode. For low values of
$\Lambda$, the EW gaugino production cross section 
is large with values in the 
hundreds of fb at $\Lambda = 35$\,TeV, but the cross section falls
off substantially as $\Lambda$ increases. As $\Lambda$ increases above
about 55\,TeV, the cross section for EW gaugino production starts to fall
below that of slepton production
(in particular $\tilde{\tau}_1^+ \tilde{\tau}_1^-$).
In a minimal model, these sleptons modes would become the dominant
SUSY production modes, but here the cross sections for slepton production
fall far below that of deltino pair production. Thus the dominant SUSY
production modes here are deltino pair production and, at values of
$\Lambda$ below 45\,TeV or so, $\chi_1^+ \chi_1^-$ and
$\chi_2^0 \, \chi_1^\pm$ production.

The decay chains depend on which sparticle is the NLSP\@. For the
values of the parameters that are considered here, either the lighter
stau or the deltino is the NLSP\@. Since the mass of the lighter stau
increases with increasing $\Lambda$, the lighter stau is the NLSP for
lower values of $\Lambda$, while the deltino is the NLSP for higher values
of $\Lambda$. For a messenger scale deltino mass of 90\,GeV, the
lighter stau is the NLSP for $\Lambda$ below about 43\,TeV\@.
For $M_{\tilde{\Delta}} (M) = 120$ and 150\,GeV, the boundaries are given
by about 58 and 73\,TeV, respectively. When the ligher stau is the NLSP,
it decays via $\tilde{\tau}_1 \rightarrow \tau \, \tilde{G}$,
and the deltino decays via the two-body mode
$\tilde{\Delta}^c \rightarrow \tilde{\tau}_1 \tau$. Then deltino pair
production leads to four $\tau$ leptons.
On the other hand, if the deltino is the NLSP,
it decays via the stau mediated three-body mode
$\tilde{\Delta}^c \rightarrow \tau \tau \tilde{G}$. So, once again, deltino
pair production again leads to the production of four $\tau$ leptons.

The decays of the lighter selectron and smuon are given in
Table~\ref{selbr}. At values of $\Lambda$ around 35 to 40\,TeV,
the neutralino is lower in mass than the $\tilde{e}_1$ and
$\tilde{\mu}_1$.  When this is the case, the main decay mode of the
is $\tilde{e}_1 \rightarrow \chi_1^0 \, e$ and the   
smuon decay is correspondingly $\tilde{\mu}_1 \rightarrow \chi_1^0 \, \mu$.
When this decay is not kinematically allowed, the decay
$\tilde{e}_1 \rightarrow \tilde{\Delta}^c \, e$ is typically
dominant if kinematically allowed. If it isn't, then the selectron
decays via the three-body
decays $\tilde{e}_1^+ \rightarrow e^+ \tau^+ \tilde{\tau}_1^-$ and
$\tilde{e}_1^+ \rightarrow e^+ \tau^- \tilde{\tau}_1^+$ and/or the
two-body mode $\tilde{e}_1 \rightarrow e \, \tilde{G}$.

The branching ratios for the neutralinos and lighter chargino are
given in Table~\ref{gaugbr}. The lighter neutralino has only the three
decay modes $\chi_1^0 \rightarrow \tilde{\tau}_1 \tau$,
$\chi_1^0 \rightarrow \tilde{\mu}_1 \mu$ and
$\chi_1^0 \rightarrow \tilde{e}_1 e$ over the parameter space considered.
Since the lighter neutralino is lighter than $\tilde{\mu}_1$ and
$\tilde{e}_1$ for $\Lambda < 43$\,TeV, the only decay mode for
$\chi_1^0$  is
$\chi_1^0 \rightarrow \tilde{\tau}_1 \tau$.
As $\Lambda$ increases beyond the point where the decays to the selectron
and smuon become kinematically available, the branching ratios for
$\chi_1^0 \rightarrow \tilde{\mu}_1 \mu$ and
$\chi_1^0 \rightarrow \tilde{e}_1 e$ increase, but the
$\chi_1^0 \rightarrow \tilde{\tau}_1 \tau$ decay remains dominant
due in large part to the fact that the mass of $\tilde{\tau}_1$ is
much lower than that of the selectron and smuon.

The chargino has only two decay modes over the allowed parameter
space: $\chi_1^\pm \rightarrow \tilde{\tau}_1 \nu_\tau$ and
$\chi_1^\pm \rightarrow \chi_1^0 \, W$. At the lower values of
$\Lambda$ considered, the decay to the lighter stau is either the
only decay mode available or is the dominant decay mode.
For $\Lambda$ around 40\,TeV and below, the only decay mode for the
lighter chargino is $\chi_1^\pm \rightarrow \tilde{\tau}_1 \, \nu_\tau$.
For these values of $\Lambda$, the lighter stau decays via
$\tilde{\tau}_1 \rightarrow  \tau \, \tilde{G}$ as discussed above.
Thus in chargino pair production, two $\tau$ leptons are produced.
As $\Lambda$ increases, the decay mode
$\chi_1^\pm \rightarrow \chi_1^0 \, W$ appears. With the subsequent
decays of the lighter neutralino to the sleptons and with the deltino as the
NLSP, the number of $\tau$ leptons produced is typically four or six
(in principle eight $\tau$ leptons can be produced although this requires
the rather rare three-body decays of the selectron and smuon).

The branching ratios of the second lightest neutralino are
particularly sensitive to the value of $\Lambda$. At lower
values of $\Lambda$, the decays to the sleptons are dominant.
In particular, the decay $\chi_2^0 \rightarrow \tilde{\tau}_1 \tau$
is dominant due to the lower mass of the $\tilde{\tau}_1$ and the
fact that the $\chi_2^0$ is mostly wino. When the $\tilde{\tau}_1$
is the NLSP, $\chi_2^0 \, \chi_1^\pm$ production typically produces
three $\tau$-jets. When the lighter stau isn't the NLSP and decays
via $\tilde{\tau}_1 \rightarrow \tau \, \tilde{\Delta}^c$, then
five $\tau$ leptons are usually produced, although three is also
common due to the decays of the $\chi_2^0$ to the $\tilde{\mu}_1$ and
$\tilde{e}_1$ followed by their decays to the deltino.
As $\Lambda$ increases, the decay
$\chi_2^0 \rightarrow \chi_1^0 \, {\rm h}$ becomes dominant, but
at these values of $\Lambda$, the cross section for EW gaugino production
falls far below that of deltino pair production.

In summary, the dominant SUSY production modes at low values of
$\Lambda$ are deltino pair production and EW gaugino production.
We expect four $\tau$ leptons to be produced in deltino pair production,
while EW gaugino production is typically expected to produce two to 
three $\tau$
leptons. For larger values of $\Lambda$, the possibility exists to
produce many $\tau$ leptons in EW gaugino production, but
the cross sections for such production are much smaller than that
for deltino pair production. Thus
four $\tau$ leptons are generally produced at larger values of $\Lambda$.

We now consider the observability of these modes at Tevatron's Run II\@.
Tables~\ref{sbr90}, \ref{sbr120} and \ref{sbr150} give the inclusive
$\tau$-jet production cross sections for a messenger scale deltino mass of
90, 120 and 150\,GeV, respectively. We include in the figures only up
to four $\tau$-jets as the cross sections for more than four $\tau$-jets 
are small. Considering Table~\ref{sbr90}, we see that before cuts the
production of
two and three $\tau$-jets are dominant, but the four $\tau$-jet cross section 
is also significant at slightly over 100\,fb.
After the cuts are applied, however, the situation
changes substantially. The one $\tau$-jet mode is now dominant, but the
cross section for two $\tau$-jets is not far below and the three 
$\tau$-jets cross section is not insignificant.

We first consider the $M_{\tilde{\Delta}} (M) = 90$\,GeV case.
We see from Table~\ref{sbr90} that for $\Lambda = 35$\,TeV the cross
section for inclusive production of three $\tau$-jets is 32.3\,fb.
For an integrated luminosity of 2\,fb$^{-1}$ (the approximate 
intial value at Run II), this corresponds to about 65 events. For 
30\,fb$^{-1}$, the number of observable events is $\sim 970$.
For $\Lambda = 85$\,TeV, the production cross section for three $\tau$-jets
has gone down slightly due to the decrease in production of charginos and
neutralinos. With a value of 26.7\,fb, the number of expected events is
about 53 and 800 for 2\,fb$^{-1}$ and 30\,fb$^{-1}$ of data, respectively.
The cross section for two $\tau$-jets is considerably higher. 
For $\Lambda = 35$\,TeV, the $\sigma \cdot {\rm BR}$ for two 
$\tau$-jets is 125\,fb which
corresponds to 250 events for 2\,fb$^{-1}$ of data and 3750 events for
30\,fb$^{-1}$ of data. For $\Lambda = 85$\,TeV, 
$\sigma \cdot {\rm BR}$ has decreased to 79\,fb. This gives about
160 and 2370 events for 2\,fb$^{-1}$ and 30\,fb$^{-1}$ of data,
respectively. 
In comparison to the GMSB model with the MSSM symmetry, 
the two $\tau$-jets and the three $\tau$-jets cross sections
are considerably higher in this model. 
In the GMSB model with MSSM symmetry, the two $\tau$-jets cross section can
be seen at RUN II, but not the three $\tau$ jets \cite{mdn}.

As the mass of the deltino increases, the production rates go down
and more variations appear. The inclusive $\tau$-jet cross sections
for $M_{\tilde{\Delta}} (M) = 120$\,GeV are shown in Table~\ref{sbr120}.
Considering the inclusive three $\tau$-jets mode, the production 
cross section
at $\Lambda = 35$\,TeV is 26\,fb. This corresponds to 52 events for
2\,fb$^{-1}$ of data and 780 events for 30\,fb$^{-1}$ of data.
This goes down to about 17\,fb at $\Lambda = 85$\,TeV\@. This gives about
34 and 510 events for 2\,fb$^{-1}$ and 30\,fb$^{-1}$ of data, 
respectively. The production rate for two $\tau$-jets is higher.
At $\Lambda = 35$\,TeV, $\sigma \cdot {\rm BR} = 94$\,fb which gives
190 and 2820 events for 2\,fb$^{-1}$ and 30\,fb$^{-1}$ of data,
respectively.

\section{Angular Distributions}

The excess of $\tau$-jets expected in this model does not constitute
an unequivocal signal for this model. $\tau$-jets are part of the
signatures for other models including the minimal GMSB model
when the lighter stau is the NLSP\@. The question then arises as to
whether there is any way to distinguish this model from the minimal
GMSB model. A possible distinguishing characteristic is the
distribution in angle between 
the two highest $E_T$ $\tau$-jets when they come from same sign $\tau$-jets.

Consider deltino pair production.
The deltino tends to decay to like sign $\tau$ leptons. This occurs
directly when the deltino is the NLSP and so decays 
via the three-body decay
$\tilde{\Delta}^{\pm \pm} \rightarrow \tau^\pm \tau^\pm \tilde{G}$.
When the two-body decay of the deltino
$\tilde{\Delta}^{\pm \pm} \rightarrow \tilde{\tau}^\pm_1 \tau^\pm$
occurs, then the second
like sign $\tau$ lepton comes from the subsequent decay of the stau.
In the rest frame of the deltino, the $\tau$ leptons are widely
distributed. In the lab frame, however, the deltinos are quite energetic
and have a large velocity, especially if their masses are small. As a
consequence of this, the decay products of the deltino tend to be collimated
in the direction in which the deltino was moving. Thus when the two most
energetic $\tau$-jets have the same sign in deltino pair production,
the angle between them tends to be smaller than when the two most energetic
$\tau$-jets have opposite sign charges.

\begin{figure}
\centering
\epsfxsize=0.6\textwidth
\epsfbox[83 234 487 546]{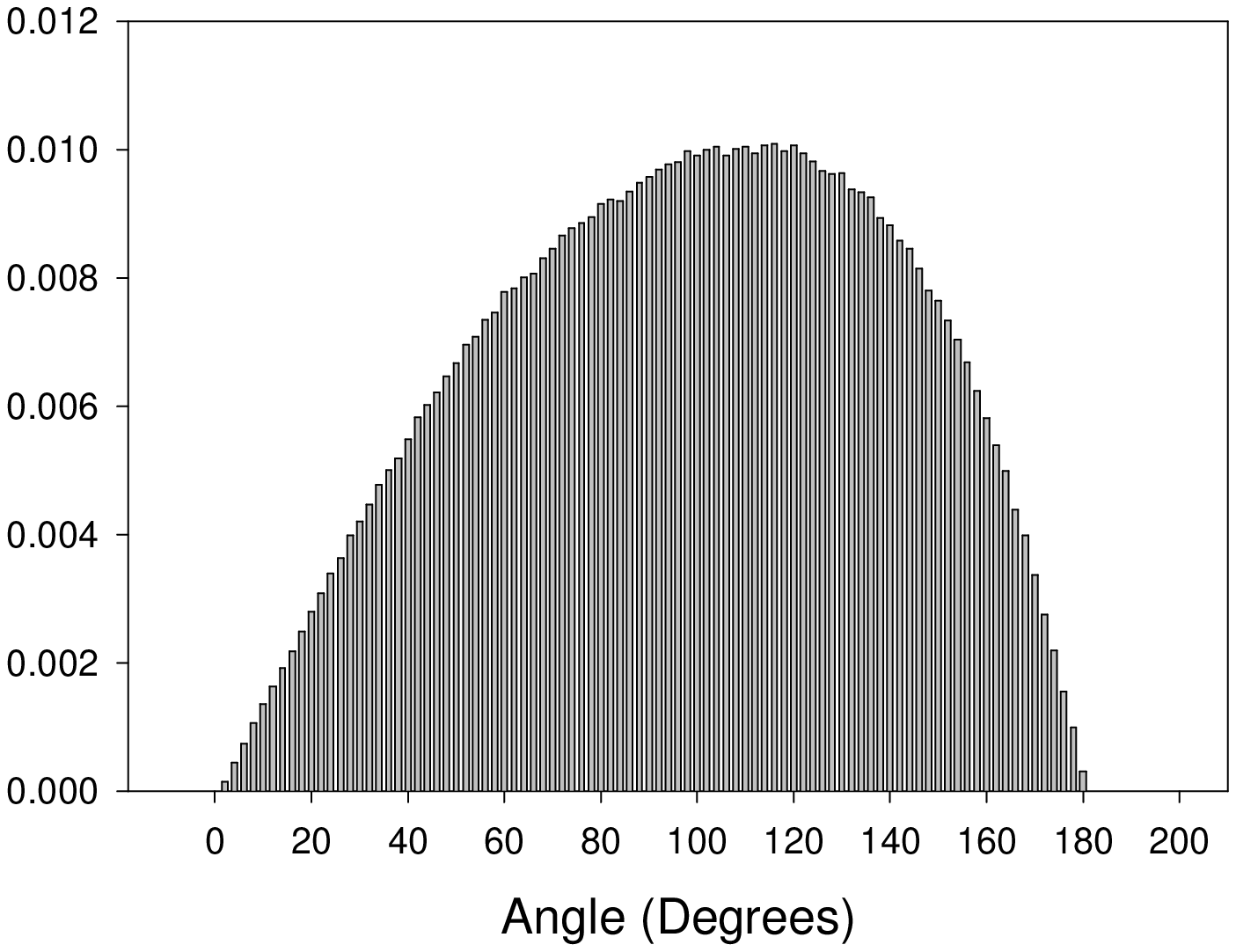}
\begin{minipage}{0.6\textwidth}
(b)
\end{minipage}
\vspace{1.5 cm}
\caption{\label{d90}
Angular distribution between the two most energetic $\tau$-jets
for deltino pair production at the Tevatron. The deltino
mass is about 97\,GeV\@.
(a) gives the distribution
when the $\tau$-jets come from same sign $\tau$ leptons. (b) gives
the distribution when the $\tau$-jets come from opposite sign $\tau$
leptons.}
\end{figure}

Fig.~\ref{d90} gives the distribution in angle between the two most
energetic $\tau$-jets for deltino pair production. This example is for
a weak scale deltino mass of about 97\,GeV\@. We can see that the
distribution in angle for like sign $\tau$-jets, which is given in
Fig.~\ref{d90}(a), peaks at about $40^\circ$. Fig.~\ref{d90}(b) gives the
distribution in angle between the two most energetic $\tau$-jets when they
come from opposite sign $\tau$ leptons. In stark contrast to the previous
case, here the peak occurs at $110^\circ$.

\begin{figure}
\centering
\epsfxsize=0.6\textwidth
\epsfbox[83 234 487 546]{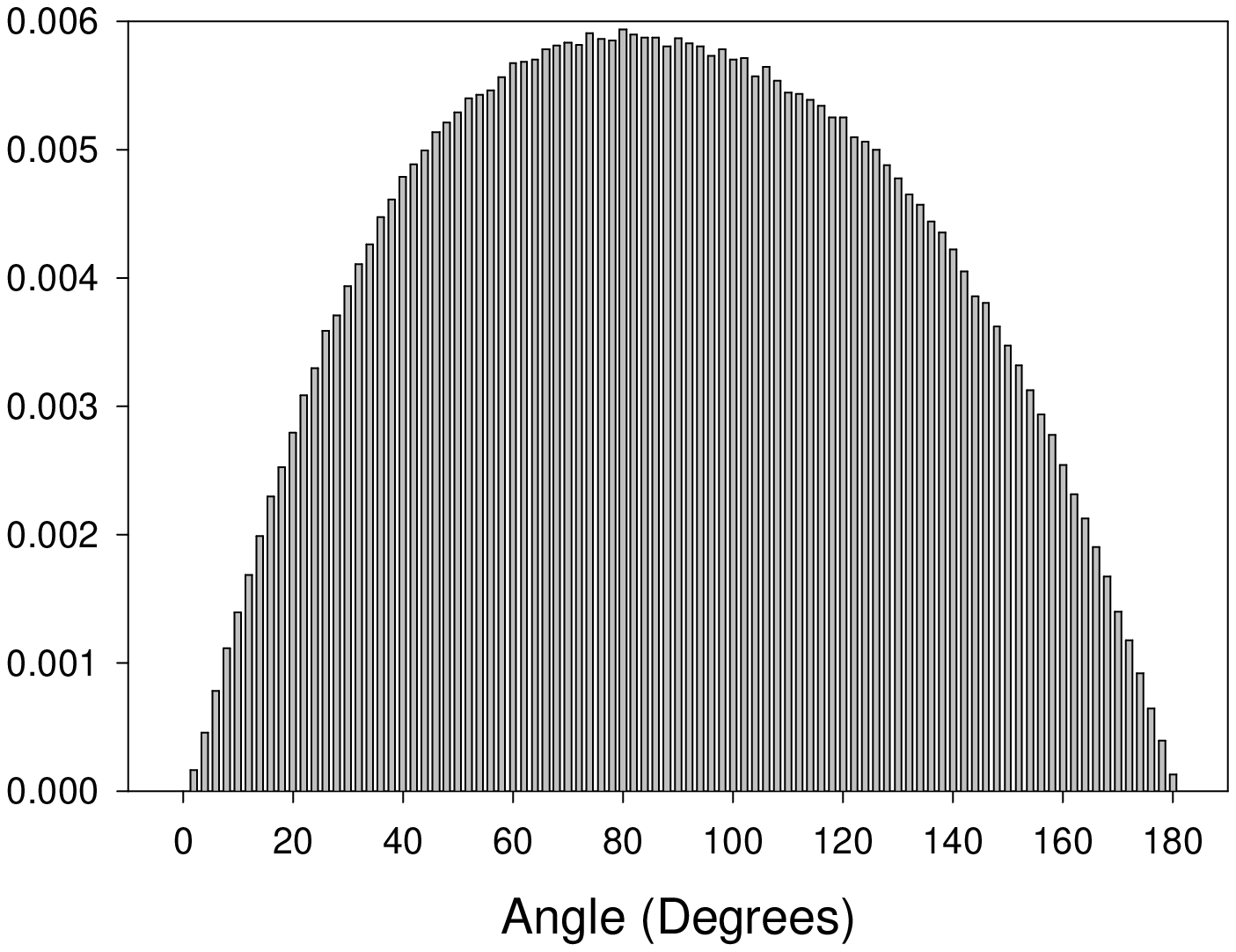}
\begin{minipage}{0.6\textwidth}
(b)
\end{minipage}
\vspace{1.5 cm}
\caption{\label{g100} Angular distribution between the two most
energetic $\tau$-jets for EW gaugino production at the Tevatron.
The $\chi_2^0$ mass is 100\,GeV\@.
(a) gives the distribution
when the $\tau$-jets come from same sign $\tau$ leptons. (b) gives
the distribution when the $\tau$-jets come from opposite sign $\tau$
leptons.}
\end{figure}

The question then arises as to how these distributions look in the usual
SUSY production modes. Fig.~\ref{g100} shows the angular distributions
for combined $\chi_2^0 \, \chi_1^\pm$ and $\chi_1^+ \chi_1^-$ production
for the input parameters $M = 100$\,TeV, $\Lambda = 45$\,TeV, $n = 1$ and
$\tan \beta = 10$. For these values of the parameters, the weak scale
$\chi_2^0$ mass is $\sim 100$\,GeV\@.
The distribution for same sign $\tau$-jets is given in Fig~\ref{g100}(a).
We see that the peak occurs at about
$110^\circ$. In this situation, same-sign $\tau$-jets do not come from
$\chi_1^+ \chi_1^-$ production. In $\chi_2^0 \, \chi_1^\pm$ production,
one of the same sign $\tau$-jets generally comes from the chargino and the
other from the neutralino.
We now consider the angular distribution for opposite sign $\tau$-jets
which are given in Fig.~\ref{g100}(b).
In $\chi_2^0 \, \chi_1^\pm$ production, opposite sign $\tau$-jets
frequently come from the neutralino, while in
$\chi_1^+ \chi_1^-$ production one of the $\tau$-jets comes from one of
the charginos and the other $\tau$-jet comes from the other chargino.
Since there is a strong possibility that the opposite sign
$\tau$-jets come from the same particle ($\chi_2^0$), the distribution
should peak at a lower angle than for same sign $\tau$-jets.
We see from the figure that the peak occurs at about $85^\circ$.

\begin{figure}
\centering
\epsfxsize=0.6\textwidth
\epsfbox[83 234 487 546]{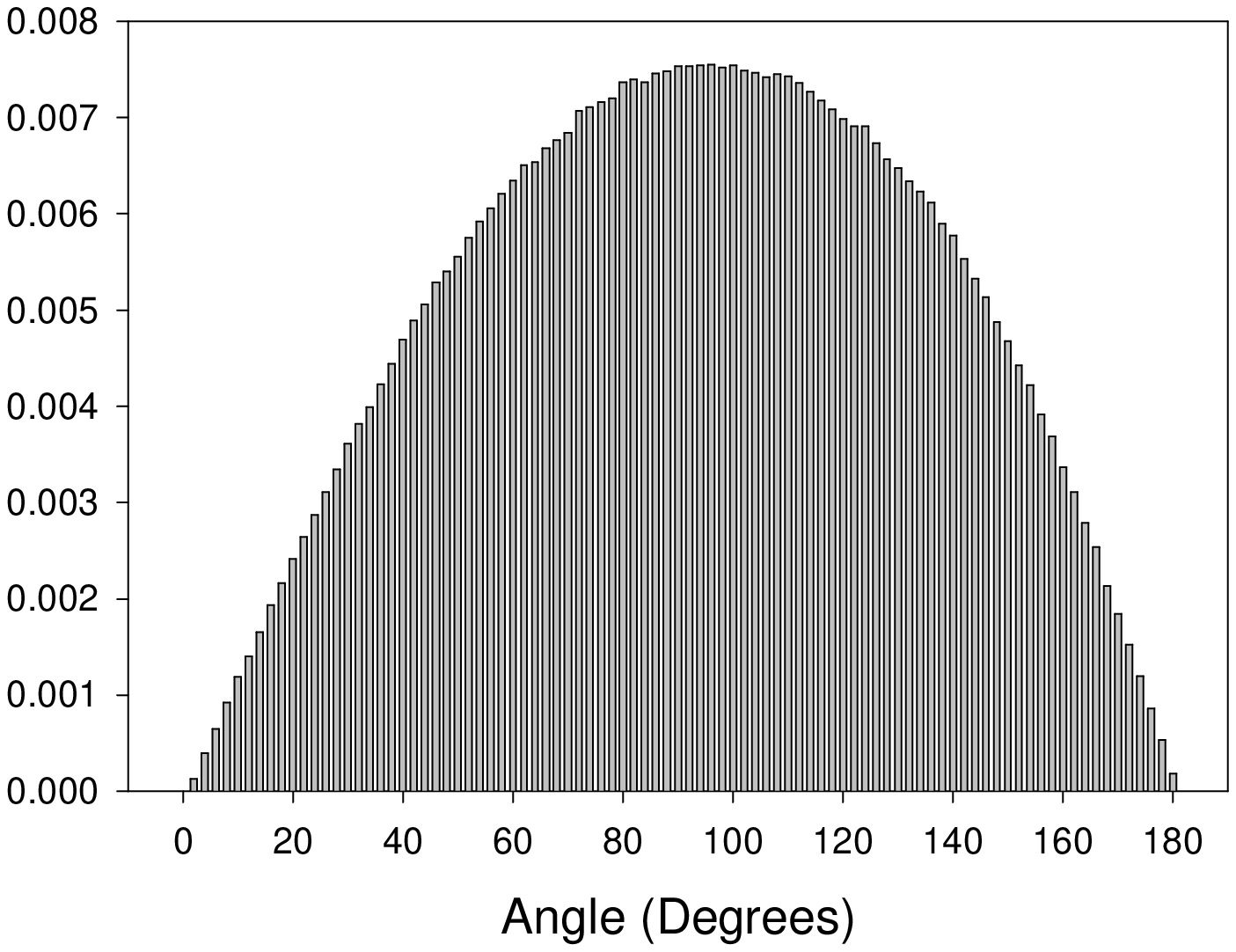}
\begin{minipage}{0.6\textwidth}
(b)
\end{minipage}
\vspace{1.5 cm}
\caption{\label{g150} Angular distribution between the two most 
energetic $\tau$-jets for EW gaugino production at the Tevatron. 
The mass of $\chi_2^0$ is about 150\,GeV\@.
(a) gives the distribution
when the $\tau$-jets come from same sign $\tau$ leptons. (b) gives
the distribution when the $\tau$-jets come from opposite sign $\tau$
leptons.}
\end{figure}

The question arises as to how much this changes as the gaugino masses
are increased. Fig.~\ref{g150} gives the angular distribution for a
$\chi_2^0$ mass of 150\,GeV\@. We see that the same sign distribution still
peaks at about $110^\circ$, while the opposite sign distribution has now
shifted to a slightly higher value of about $95^\circ$.

The actual angular distribution between the two highest $E_T$ $\tau$-jets 
depends on which SUSY production modes are
important. For certain regions of the parameter space (depending, 
in particular, on the values of $\Lambda$ and the messenger scale 
deltino mass), 
deltino pair production is the only important SUSY production mode.
When this is the case, the angular
distributions are simply given by those for deltino pair production.

\begin{figure}
\centering
\epsfxsize=0.6\textwidth
\epsfbox[83 234 487 546]{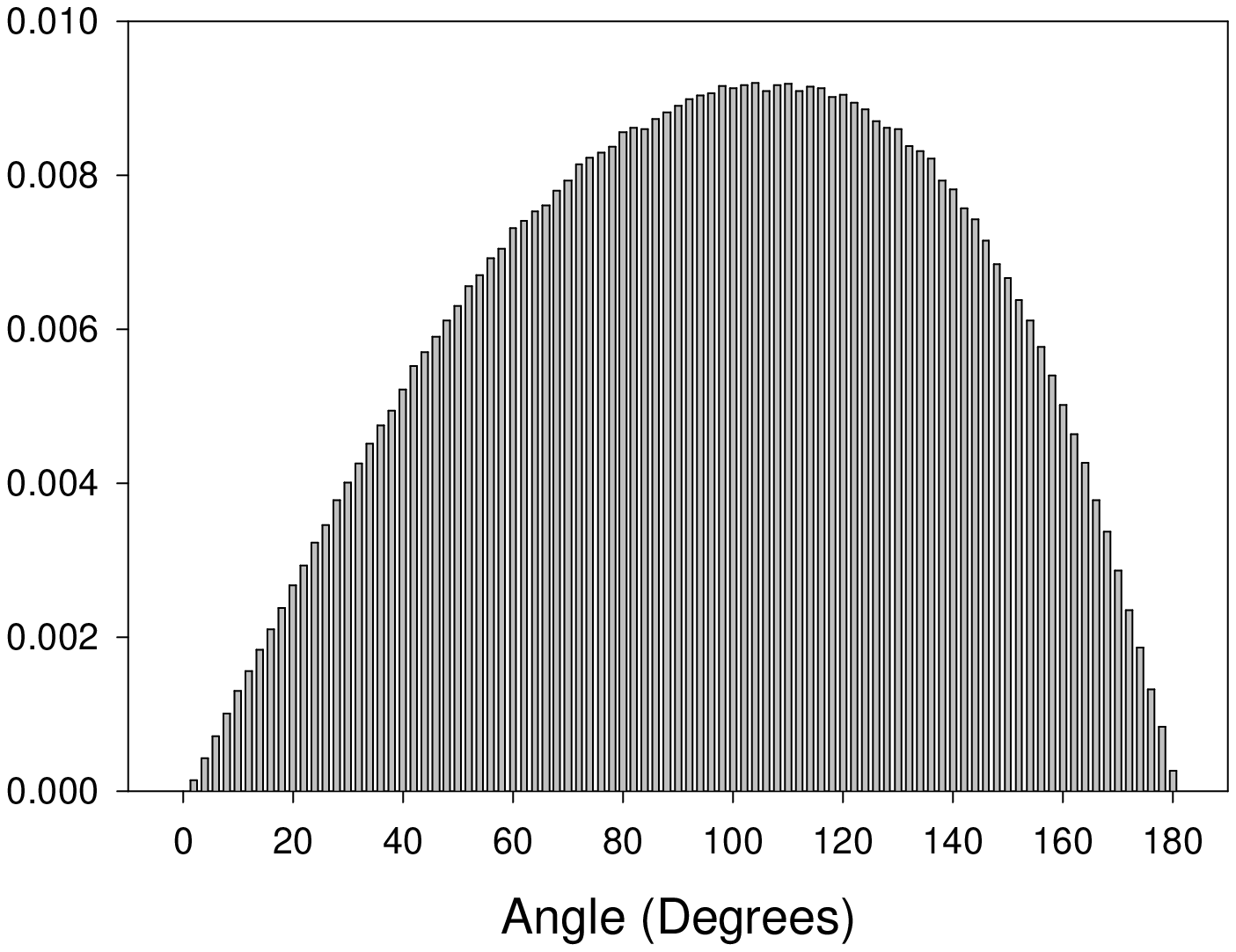}
\begin{minipage}{0.6\textwidth}
(b)
\end{minipage}
\vspace{1.5 cm}
\caption{\label{angle90} Angular distribution between the two most
energetic $\tau$-jets for combined SUSY pair production at the Tevatron.
The messenger scale deltino mass is 90\,GeV\@. The other parameters are
$\tan \beta = 15$, $n = 2$ and $M$/$\Lambda$ = 3.
(a) gives the distribution
when the $\tau$-jets come from same sign $\tau$ leptons. (b) gives
the distribution when the $\tau$-jets come from opposite sign $\tau$
leptons.}
\end{figure}

In other regions of the parameter space, 
EW gaugino production can be significant. 
We consider the angular distributions for an example given by the
input parameters $\tan \beta = 15$, $n = 2$ and 
$M$/$\Lambda$ = 3. Three values of the messenger scale deltino mass
are considered: 90, 120 and 150\,GeV\@. The angular distributions
for $M_{\tilde{\Delta}} (M) = 90$\,GeV are given in Fig.~\ref{angle90}.
Since the deltino is especially light at 96\,GeV, deltino pair production
is the dominant SUSY production mode. Thus deltino pair production
largely dictates the form of the angular distributions.

Fig.~\ref{angle90}(a) gives the angular distribution between the two
highest $E_T$ $\tau$-jets when they come from same sign $\tau$ leptons.
In the figure we can see the rather striking peak at around $40^\circ$.
This is due to the same sign $\tau$-jets coming mostly from the decay of the
same deltino. Since the deltino mass is especially light compared to the
beam energy, they typically move rapidly in the lab frame. Thus their
decay products tend to be more tightly collimated than in the production
of the heavier particles.

Fig.~\ref{angle90}(b) gives the angular distribution between the two
highest $E_T$ $\tau$-jets when they come from opposite sign
$\tau$ leptons. We see from this figure that the peak occurs at about
$110^\circ$. Here the $\tau$-jets typically come 
from the decay chains of different
particles and so the angle between the $\tau$-jets is typically quite
large.

\begin{figure}
\centering
\epsfxsize=0.6\textwidth
\epsfbox[83 234 487 546]{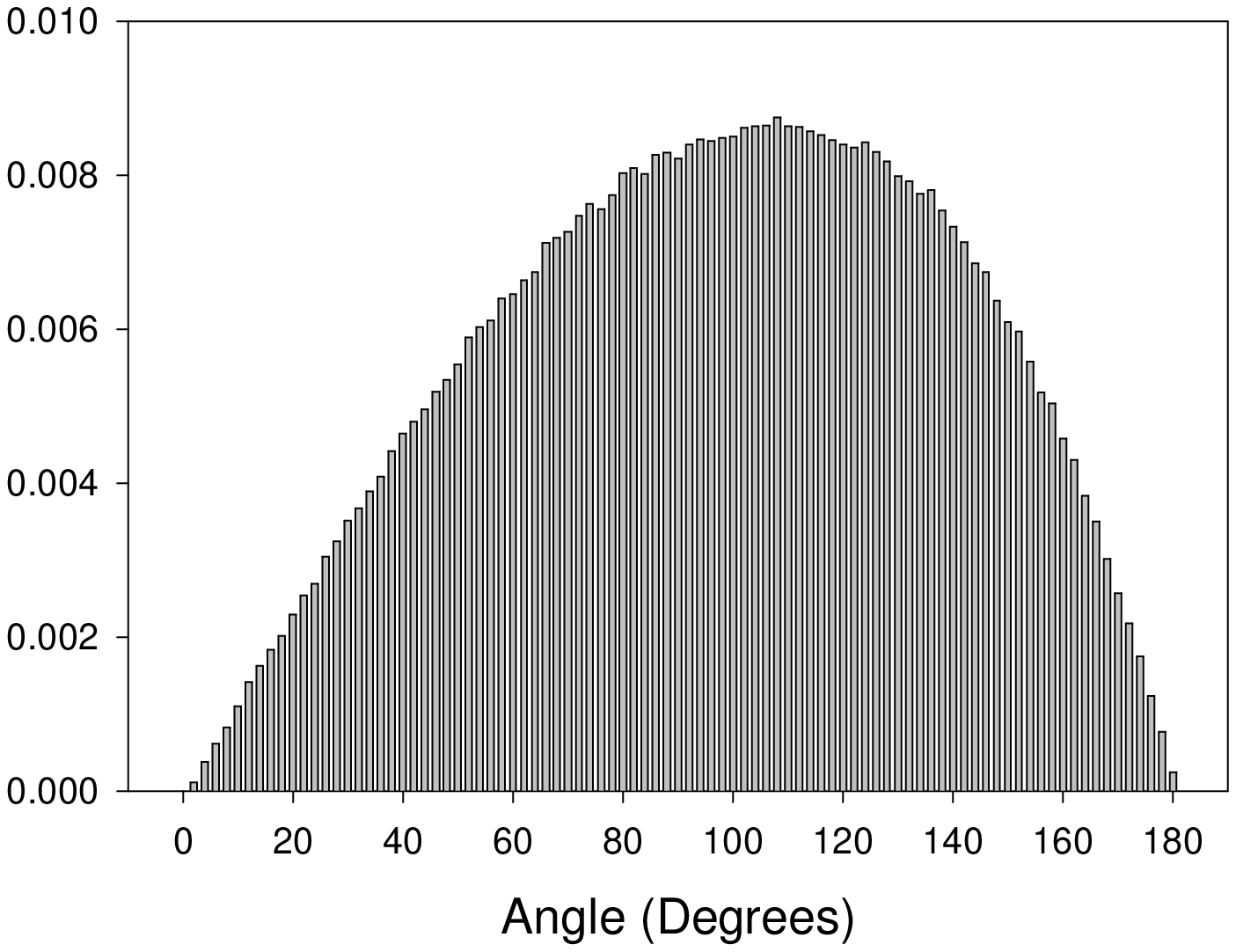}
\begin{minipage}{0.6\textwidth}
(b)
\end{minipage}
\vspace{1.5 cm}
\caption{\label{angle120} Angular distribution between the two most
energetic $\tau$-jets for combined SUSY production at the Tevatron.
The messenger scale deltino mass is 120\,GeV\@. The other parameters
are $\tan \beta = 15$, $n = 2$ and $M$/$\Lambda$ = 3.
(a) gives the distribution
when the $\tau$-jets come from same sign $\tau$ leptons. (b) gives
the distribution when the $\tau$-jets come from opposite sign $\tau$
leptons.}
\end{figure}

The situation changes as the deltino mass gets larger. This is in part
due to the fact that the deltino pair production cross section gets
smaller and so production of charginos and neutralinos can have a
larger impact on the distributions. In addition, a larger deltino mass
means the deltinos will typically be moving slower. Thus the boost effect
won't have as dramatic an effect on the deltino's decay products.
The example with $M_{\tilde{\Delta}} (M) = 120$\,GeV is given in
Fig.~\ref{angle120}. We can see that the distribution for same sign
$\tau$-jets peaks at about $70^\circ$. On the other hand, the opposite
sign $\tau$-jet angular distribution still peaks at around $110^\circ$. Thus
the angle between the $\tau$-jets is less striking a signature than
it was before, but it is still distinctive.

\begin{figure}
\centering
\epsfxsize=0.6\textwidth
\epsfbox[83 234 487 546]{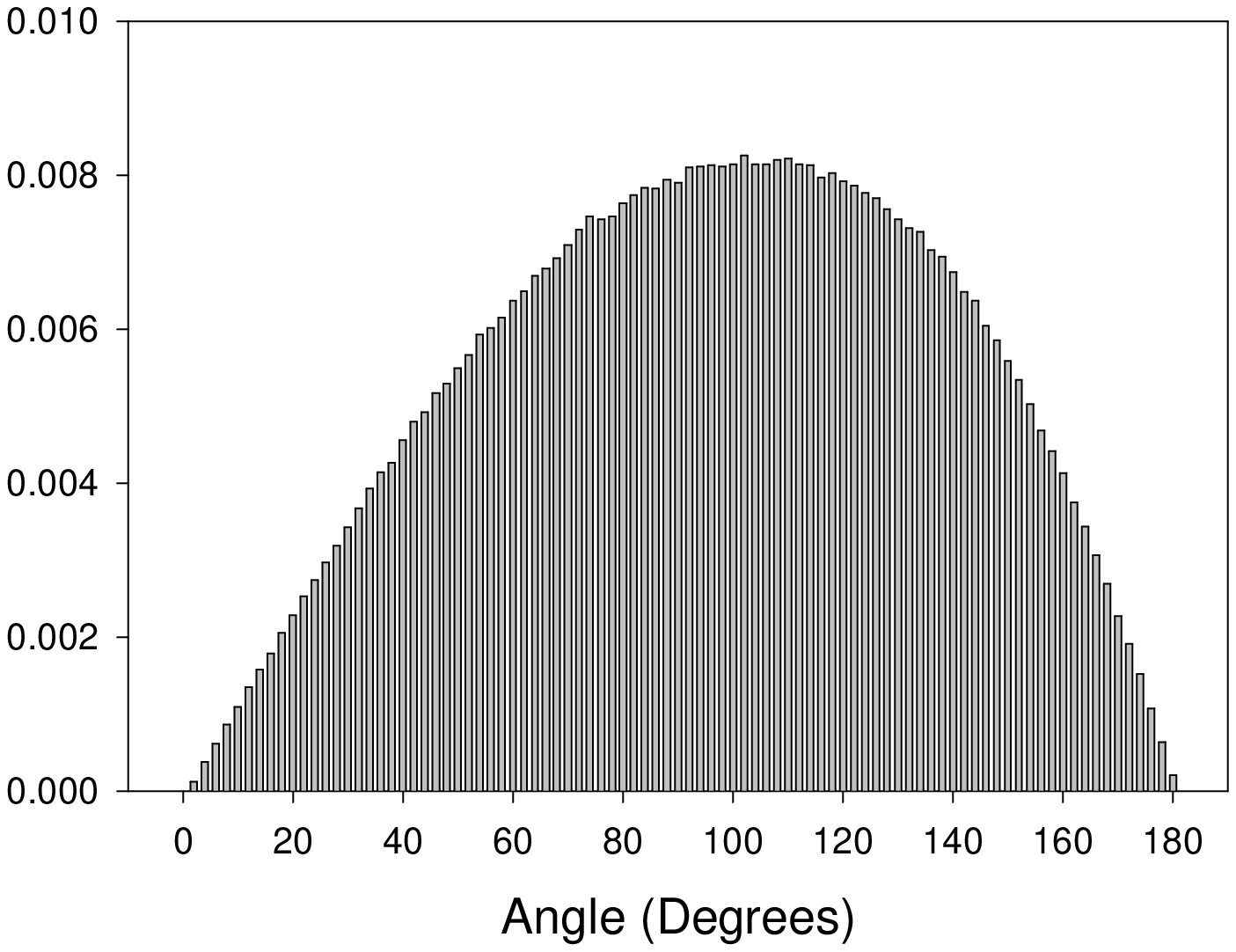}
\begin{minipage}{0.6\textwidth}
(b)
\end{minipage}
\vspace{1.5 cm}
\caption{\label{angle150} Angular distribution between the two most
energetic $\tau$-jets for combined SUSY production at the Tevatron.
The messenger scale deltino mass is 150\,GeV\@. The other parameters
are $\tan \beta = 15$, $n = 2$ and $M$/$\Lambda$ = 3.
(a) gives the distribution
when the $\tau$-jets come from same sign $\tau$ leptons. (b) gives
the distribution when the $\tau$-jets come from opposite sign $\tau$
leptons.}
\end{figure}

The results for a messenger scale deltino mass of 150\,GeV is given
in Fig.~\ref{angle150}. The peak in the distribution in angle between
the two highest $E_T$ $\tau$-jets when they have the same sign peaks
at a rather high $95^\circ$. As before, the peak in the distribution
for the two highest $E_T$ $\tau$-jets when they have opposite sign
is at $110^\circ$. Thus the distinctiveness due to the
angle between the two highest $E_T$ $\tau$-jets is nearly lost for
such a large value of the deltino mass. This is partially due to the
deltino pair cross section of 92\,fb being quite a bit lower than the
cross section for $\chi_1^\pm \, \chi_2^0$ production and
$\chi_1^+ \chi_1^-$ production. In addition, as discussed above, there
is also the reduction in the boost effect as the mass of the decaying
deltino increases.

\section{Conclusion} 

In conclusion, we have found that the doubly charged Higgs bosons of LR
models can be potentially observable at Run II of the Tevatron through 
the production of $\tau$-jets. In a
GMSB type theory, SUSYLR models typically produce large numbers of
two and three $\tau$-jet final states. 
This large $\tau$-jet signal is 
also due in large part to pair production of the doubly charged 
Higgsino.
It is also due to the relatively low
mass of the lighter stau (which is frequently the NLSP) in these models,
which is due to the additional coupling $f$. 
We have also shown that the distribution in angle between the two highest
$E_T$ $\tau$-jets is different from other models which do not have this
doubly charged Higgsino.

\section*{Acknowledgments}

The work of R. N. M. is supported by National Science Foundation grant No.
PHY-9421385 and that of D.~M. by DE-FG03-98ER41076. We thank  T.~Kamon for
 valuable  comments and suggestions.

\newpage

\clearpage

\begin{table}
\centering
\caption{\label{delcross}  Cross sections (in fb) for deltino pair production
for various values of $\Lambda$. The other parameters used are
$\tan \beta = 15$, $n = 2$ and $M$/$\Lambda$ = 3.}
\vskip 0.25cm
\begin{tabular}{l c c c}
$M_{\tilde{\Delta}} (M)$   &   $\Lambda = 35$\,TeV  &  $\Lambda = 60$\,TeV
    &  $\Lambda = 85$\,TeV  \\
\hline
90\,GeV    &   643.0  &  629.9  &  621.5  \\
120\,GeV   &   228.2  &  222.9  &  219.6  \\
150\,GeV   &   91.6   &  89.2   &  87.7   \\
\end{tabular}
\end{table}

\begin{table}
\caption{\label{selbr} Branching ratios of the sleptons. The values of the
parameters are $\tan \beta = 15$, $n = 2$ and $M$/$\Lambda$ = 3.}
\vskip 0.25cm
\begin{tabular}{l c c c c c c c}
$\Lambda$ (TeV)  &  35  &  40  &  50  &  60  &  70  &  80  &  85  \\
\hline
\hline
\multicolumn{8}{c}{$M_{\tilde{\Delta}} (M) = 90$\,GeV}  \\
\hline
$\tilde{e}_1 \rightarrow e \, \chi_1^0$ & 1 & 0.9252  &  -  &  -  &  -
    &  -  &  -  \\
$\tilde{e}_1 \rightarrow e \, \tilde{\Delta}$  &  -  &  0.0748  &  1
    &  1  &  1  &  1  &  1  \\
\hline
$\tilde{\tau}_1 \rightarrow \tau \, \tilde{G}$  &  1  &  1  &  -  &  -
    &  -  &  -  &  -  \\
$\tilde{\tau}_1 \rightarrow \tau \, \tilde{\Delta}$  &  -  &  -  &  1
    &  1  &  1  &  1  &  1  \\
\hline
\hline
\multicolumn{8}{c}{$M_{\tilde{\Delta}} (M) = 120$\,GeV}   \\
\hline
$\tilde{e}_1 \rightarrow e \, \chi_1^0$  &  1  &  1  &  -  &  -  &  -
    &  -  &  -  \\
$\tilde{e}_1^+ \rightarrow e^+ \tau^+ \tilde{\tau}_1^-$  &  -  &  -
    &  0.2065  &  -  &  -  &  -  &  - \\
$\tilde{e}_1^+ \rightarrow e^+ \tau^- \tilde{\tau}_1^+$  &  -  &  -
    &  0.1684  &  -  &  -  &  -  &  -  \\
$\tilde{e}_1 \rightarrow e \, \tilde{\Delta}$  &  -  &  -  &  0.6251
    &  1  &  1  &  1  &  1  \\
\hline
$\tilde{\tau}_1 \rightarrow \tau \, \tilde{G}$  &  1  &  1  &  1  &  1
    &  -  &  -  &  -  \\
$\tilde{\tau}_1 \rightarrow \tau \, \tilde{\Delta}$  &  -  &  -  &  -  &  -
    &  1  &  1  &  1  \\
\hline
\hline
\multicolumn{8}{c}{$M_{\tilde{\Delta}} (M) = 150$\,GeV}     \\
\hline
$\tilde{e}_1 \rightarrow e \, \chi_1^0$  &  1  &  1  &  -  &  -  &  -
    &  -  &  -  \\
$\tilde{e}_1^+ \rightarrow e^+ \tau^+ \tilde{\tau}_1^-$  &  -  &  -
    &  0.5532  &  0.5643  &  -  &  -  &  -  \\
$\tilde{e}_1^+ \rightarrow e^+ \tau^- \tilde{\tau}_1^+$  &  -  &  -
    &  0.4468  &  0.4357  &  -  &  -  &  -  \\
$\tilde{e}_1 \rightarrow e \, \tilde{\Delta}$  &  -  &  -  &  -  &  -  &
       1  &  1  &  1  \\
\hline
$\tilde{\tau}_1 \rightarrow \tau \, \tilde{G}$  &  1  &  1  &  1  &  1
    &  1  &  -  &  -  \\
$\tilde{\tau}_1 \rightarrow \tau \, \tilde{\Delta}$  &  -  &  -  &  -
    &  -  &  -  &  1  &  1  \\
\end{tabular}
\end{table}

\begin{table}
\caption{\label{gaugbr} Branching ratios of some of the sparticles of
interest. The values of the parameters are $\tan \beta = 15$, $n = 2$ and
$M$/$\Lambda$ = 3.
The messenger scale deltino mass is 90\,GeV, but the branching ratios of
these sparticles have little dependence on the deltino mass.}
\vskip 0.25cm
\begin{tabular}{l c c c c c c c}
$\Lambda$ (TeV)  &  35  &  40  &  50  &  60  &  70  &  80  &  85  \\
\hline
$\chi_1^\pm \rightarrow \tilde{\tau}_1 \nu_\tau$  &  1  &  1  &  0.7153
        &  0.5807  &  0.5166  &  0.4796  &  0.4663  \\
$\chi_1^\pm \rightarrow \chi_1^0 \, W$  &  -  &  -  &  0.2847  &  0.4193
        &  0.4834  &  0.5204  &  0.5337  \\
\hline
$\chi_2^0 \rightarrow \tilde{\tau}_1 \tau$  &  0.6312
    &  0.6587  &  0.6860  &  0.4034  &  0.3133  &  0.2706  &  0.2560  \\
$\chi_2^0 \rightarrow \tilde{e}_1 e$  &  0.1844  &  0.1707
    &  0.1366  &  0.0625  &  0.0386  &  0.0270  &  0.0232  \\
$\chi_2^0 \rightarrow \tilde{\mu}_1 \mu$  &  0.1844  &  0.1707
    &  0.1366  &  0.0625  &  0.0386  &  0.0270  &  0.0232  \\
$\chi_2^0 \rightarrow \chi_1^0 \, Z$  &  -  &  -  &  0.0408  &  0.0325
    &  0.0277  &  0.0252  &  0.0243  \\
$\chi_2^0 \rightarrow \chi_1^0 \, {\rm h}$  &  -  &  -  &  -  &  0.4392
    &  0.5818  &  0.6501  &  0.6733  \\
\hline
$\chi_1^0 \rightarrow \tilde{\tau}_1 \tau$  &  1  &  1
    &  0.9692  &  0.9099  &  0.8723  &  0.8506  &  0.8436  \\
$\chi_1^0 \rightarrow \tilde{e}_1 e$  &  -  &  -  &  0.0154
    &  0.0451  &  0.0638  &  0.0747  &  0.0782  \\
$\chi_1^0 \rightarrow \tilde{\mu}_1 \mu$  &  -  &  -  &  0.0154
    &  0.0451  &  0.0638  &  0.0747  &  0.0782  \\
\end{tabular}
\end{table}

\begin{table}
\caption{\label{sbr90} Inclusive $\tau$-jet production cross sections for
a messenger scale deltino mass of 90\,GeV\@. The other parameters are
$\tan \beta = 15$, $n = 2$ and $M$/$\Lambda$ = 3.}
\vskip 0.25cm
\begin{tabular}{l c c c c c c c}
$\Lambda$ (TeV)  &  35  &  40  &  50  &  60  &  70  &  80  &  85  \\
\cline{2-8}
    &  \multicolumn{7}{c}{$\sigma \cdot {\rm BR}$ (fb)}  \\
\hline
1 $\tau$-jet: before cuts  &  198.6  &  132.9  &  73.11  &  71.64  &  70.86
             &  70.34  &  70.14  \\
after cuts   &  168.7  &  156.2  &  97.49  &  91.90  &  90.15  &  89.33
             &  89.05  \\
\hline
2 $\tau$-jets: before cuts &  362.6  &  277.3  &  203.6  &  196.7  &  196.3
             &  194.8  &  194.2  \\
after cuts   &  124.6  &  93.11  &  89.59  &  82.91  &  80.59  &  79.63
             &  79.33  \\
\hline
3 $\tau$-jets: before cuts &  306.0  &  273.5  &  253.6  &  244.7  &  240.9
             &  238.8  &  238.0  \\
after cuts   &  32.31  &  17.75  &  32.45  &  29.02  &  27.53  &  26.88
             &  26.73  \\
\hline
4 $\tau$-jets: before cuts &  119.4  &  116.9  &  126.42  &  116.5  &  113.1
             &  111.6  &  111.1  \\
after cuts   &  3.21  &  1.18  &  5.26  &  4.28  &  3.67  &  3.45  &  3.39  \\
\end{tabular}
\end{table}

\begin{table}
\caption{\label{sbr120} Inclusive $\tau$-jet production cross sections for
a messenger scale deltino mass of 120\,GeV\@. The other parameters are 
$\tan \beta = 15$, $n = 2$ and $M$/$\Lambda$ = 3.}
\vskip 0.25cm
\begin{tabular}{l c c c c c c c}
$\Lambda$ (TeV)  &  35  &  40  &  50  &  60  &  70  &  80  &  85  \\
\cline{2-8}
    &  \multicolumn{7}{c}{$\sigma \cdot {\rm BR}$ (fb)}  \\
\hline
1 $\tau$-jet: before cuts  &  151.5  &  86.21  &  39.60  &  29.72  &  25.30
             &  25.00  &  24.90  \\
after cuts   &  125.3  &  85.20  &  53.68  &  75.87  &  42.16  &  41.38
             &  41.16  \\
\hline
2 $\tau$-jets: before cuts &  232.8  &  148.0  &  90.27  &  75.93  &  70.06
             &  69.15  &  68.83  \\
after cuts   &  93.94  &  71.59  &  49.79  &  28.57  &  43.81  &  42.83
             &  42.55  \\
\hline
3 $\tau$-jets: before cuts &  147.6  &  115.9  &  96.70  &  88.92  &  86.55
             &  85.16  &  84.71  \\
after cuts   &  25.99  &  23.22  &  17.33  &  0.55  &  17.37  &  16.80
             &  16.64  \\
\hline
4 $\tau$-jets: before cuts &  46.22  &  44.03  &  42.59  &  40.70  &  40.72
             &  39.55  &  39.22  \\
after cuts   &  3.05  &  3.09  &  2.29  &  0.06  &  2.70  &  2.50  &  2.44  \\
\end{tabular}
\end{table}

\begin{table}
\caption{\label{sbr150} Inclusive $\tau$-jet production cross sections for
a messenger scale deltino mass of 150\,GeV\@. The other parameters are
$\tan \beta = 15$, $n = 2$ and $M$/$\Lambda$ = 3.}
\vskip 0.25cm
\begin{tabular}{l c c c c c c c}
$\Lambda$ (TeV)  &  35  &  40  &  50  &  60  &  70  &  80  &  85  \\
\cline{2-8}
             &  \multicolumn{7}{c}{$\sigma \cdot {\rm BR}$ (fb)}  \\
\hline
1 $\tau$-jet: before cuts  &  136.2  &  70.92  &  24.42  &  14.61  &  11.80
             &  10.07  &  10.00  \\
after cuts   &  105.3  &  65.54  &  32.12  &  22.59  &  23.50  &  19.28
             &  18.97  \\
\hline
2 $\tau$-jets: before cuts &  190.3  &  105.7  &  48.32  &  34.22  &  30.01
             &  27.89  &  27.67  \\
after cuts   &  72.17  &  50.55  &  31.35  &  24.29  &  18.99  &  21.75
             &  21.46  \\
\hline
3 $\tau$-jets: before cuts &  95.31  &  63.85  &  45.21  &  37.72  &  35.14
             &  34.47  &  34.14  \\
after cuts   &  17.23  &  14.64  &  12.24  &  10.14  &  5.53  &  9.32
             &  9.24  \\
\hline
4 $\tau$-jets: before cuts &  21.81  &  19.75  &  18.58  &  17.00  &  16.16
             &  16.23  &  15.96  \\
after cuts   &  1.73  &  1.81  &  1.89  &  1.67  &  0.59  &  1.46  &  1.47  \\
\end{tabular}
\end{table}

\clearpage

\begin{figure}
\centering
\epsfxsize=0.6\textwidth
\epsfbox[82 234 488 544]{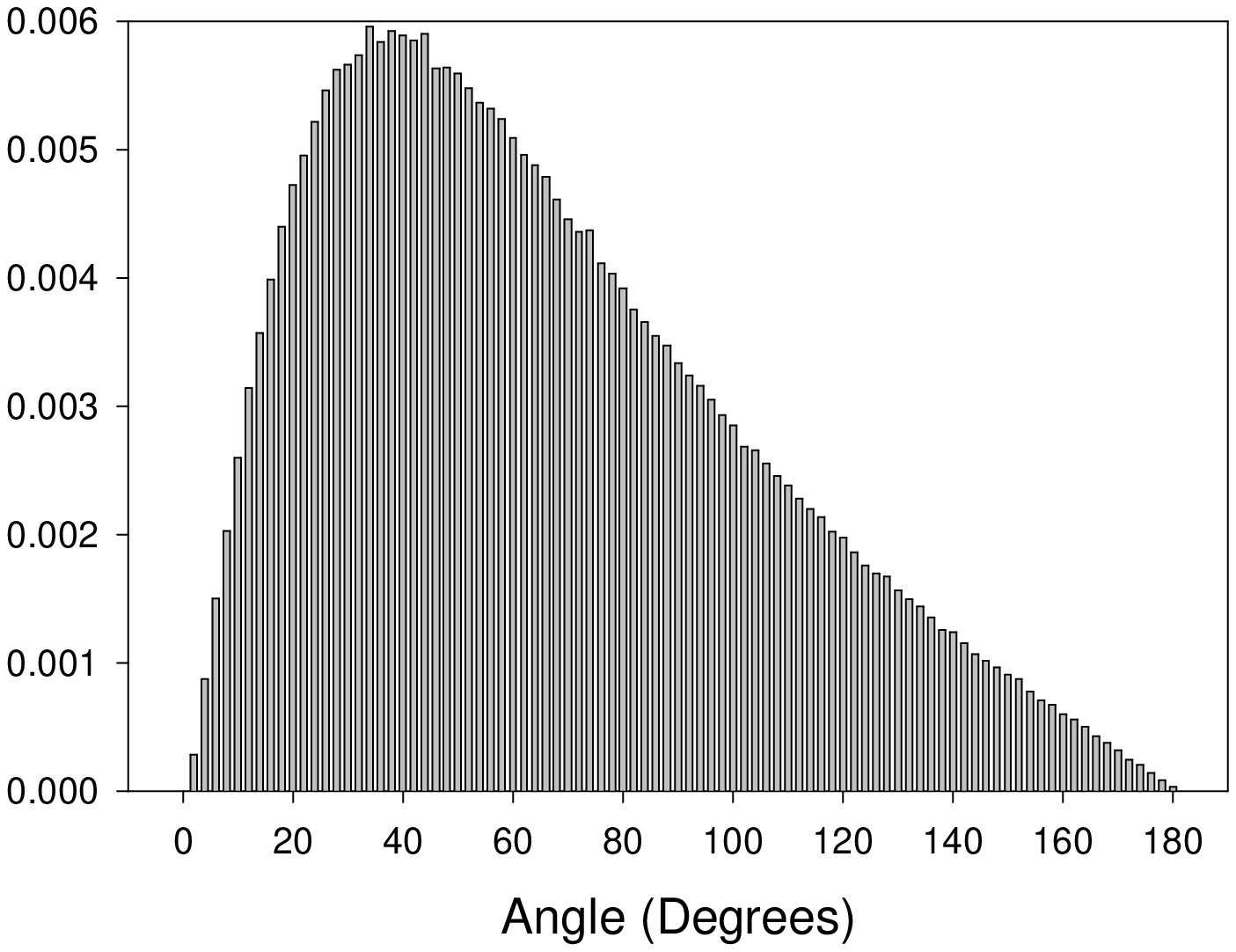}
\begin{minipage}{0.6\textwidth}
(a)
\end{minipage}
\end{figure}

\vspace{0.5 cm}

\begin{figure}
\centering
\epsfxsize=0.6\textwidth
\epsfbox[76 234 488 544]{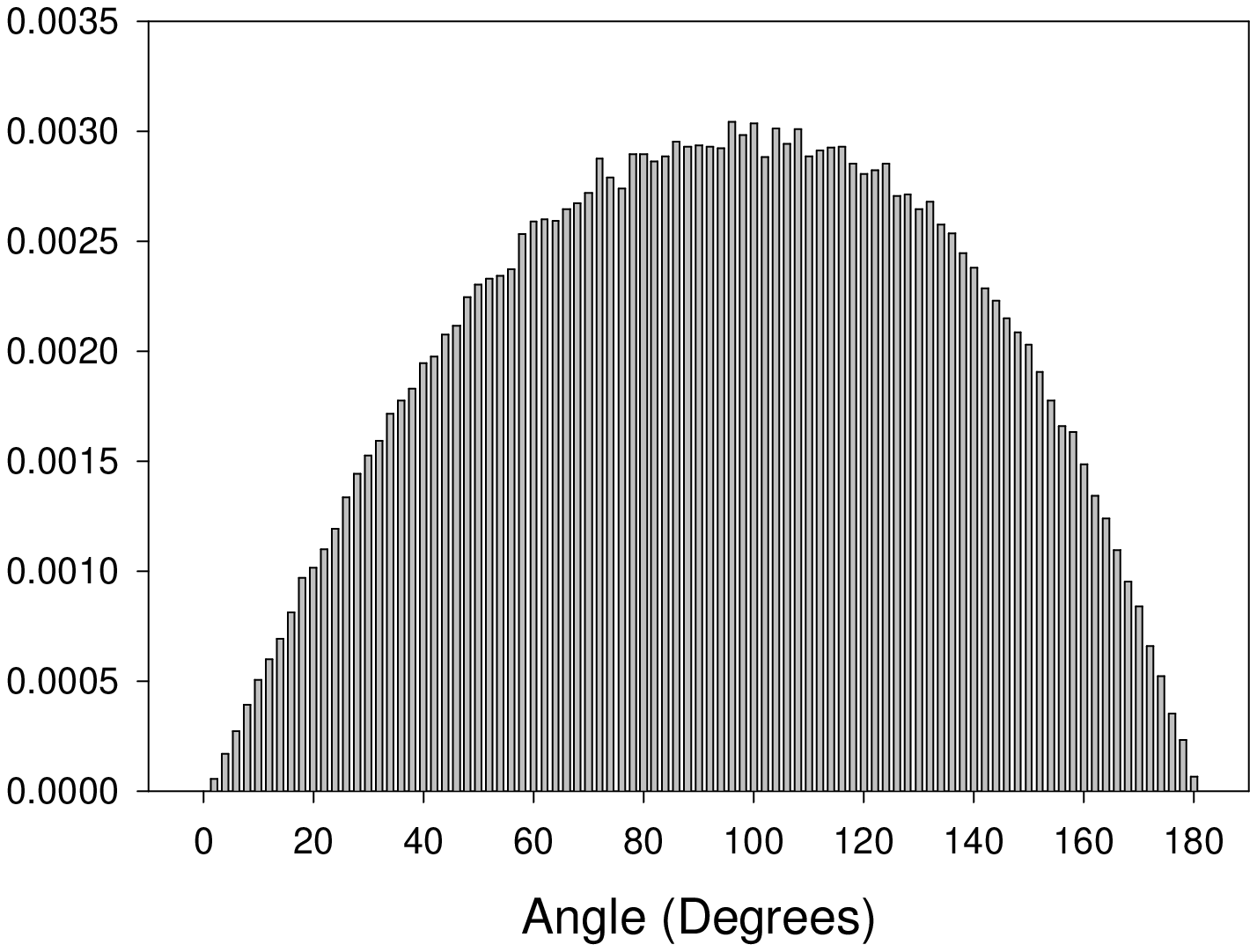}
\begin{minipage}{0.6\textwidth}
(a)
\end{minipage}
\end{figure}

\vspace{0.5 cm}

\begin{figure}
\centering
\epsfxsize=0.6\textwidth
\epsfbox[82 234 488 544]{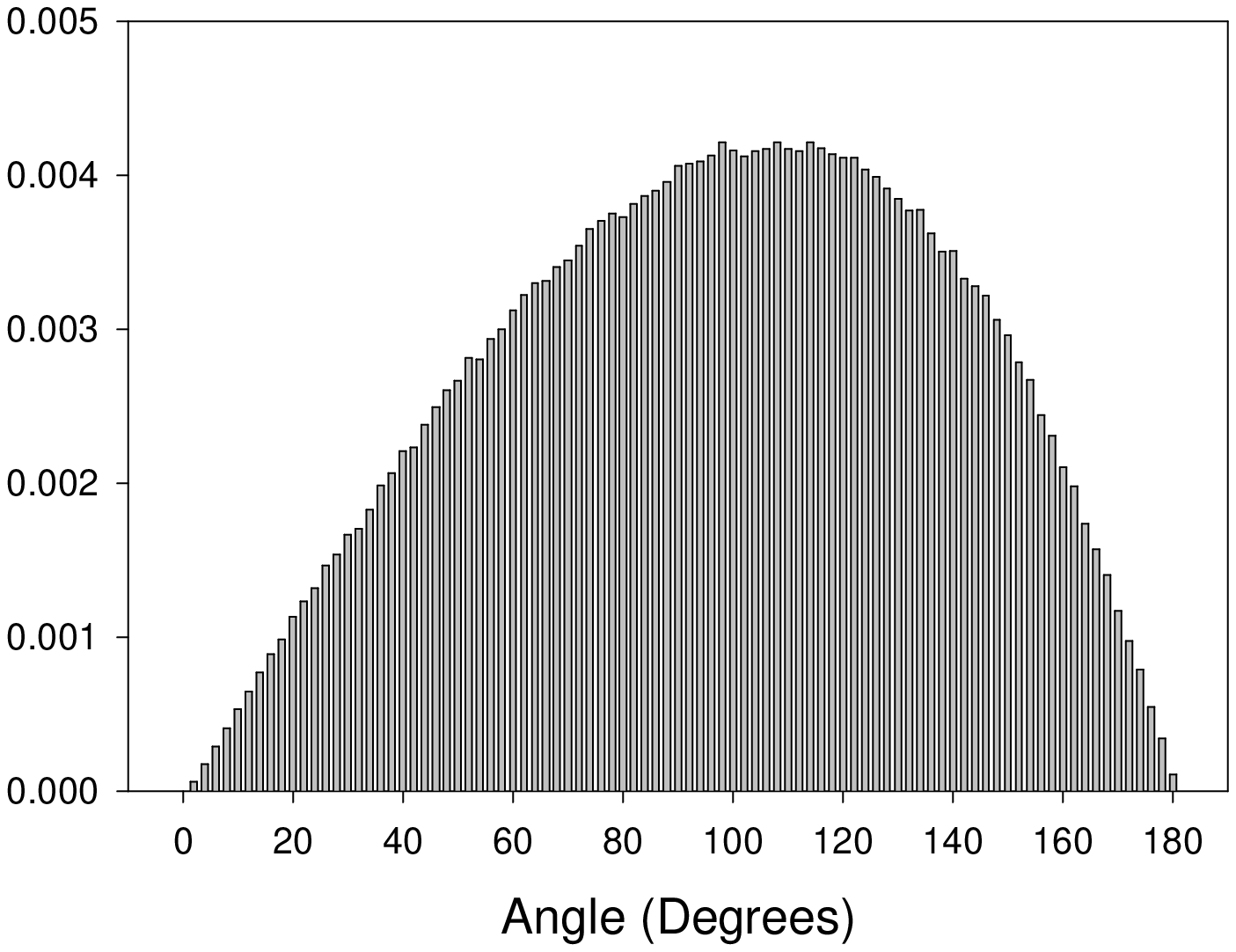}
\begin{minipage}{0.6\textwidth}
(a)
\end{minipage}
\end{figure}

\vspace{0.5 cm}

\begin{figure}
\centering
\epsfxsize=0.6\textwidth
\epsfbox[82 234 488 544]{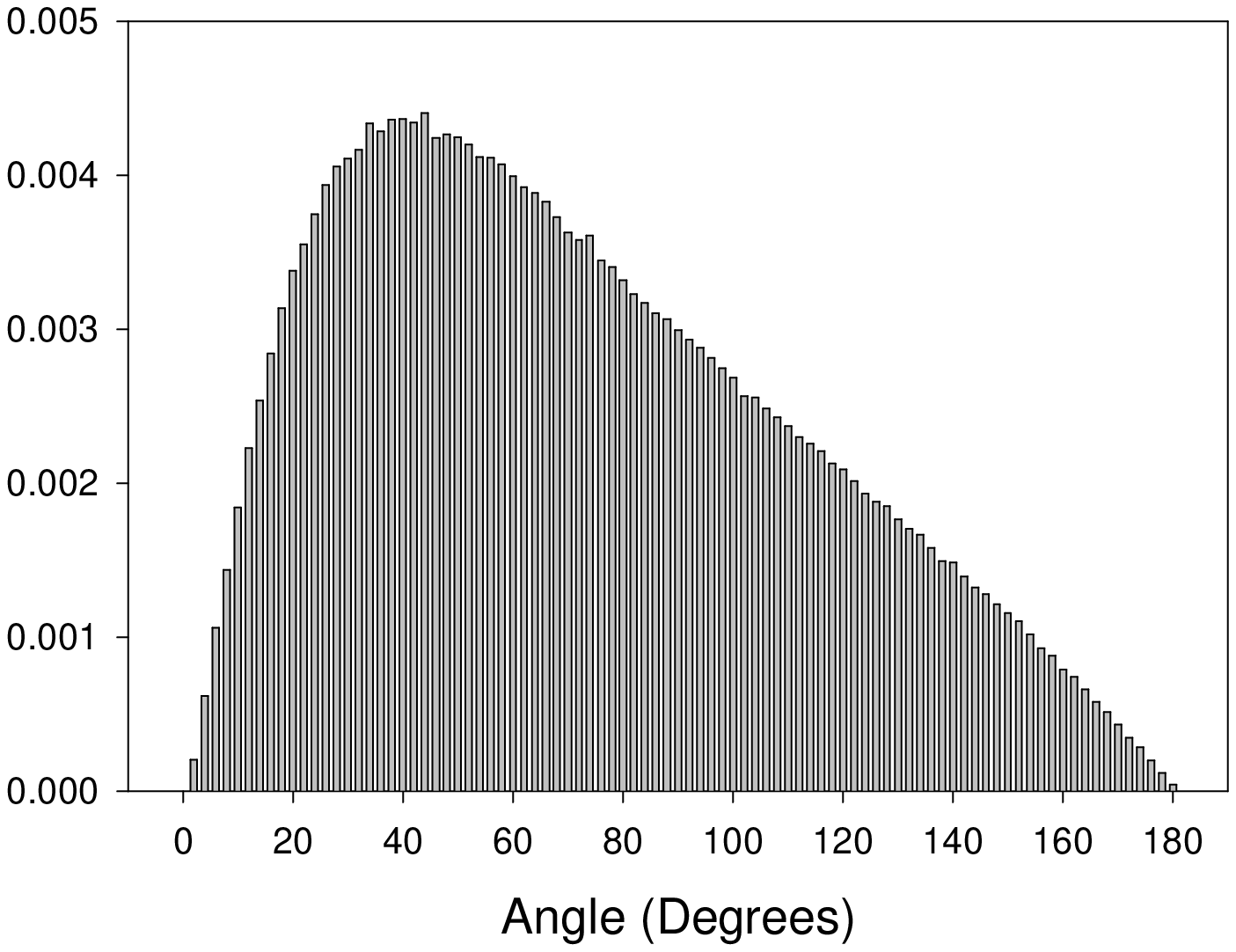}
\begin{minipage}{0.6\textwidth}
(a)
\end{minipage}
\end{figure}

\vspace{0.5 cm}

\begin{figure}
\centering
\epsfxsize=0.6\textwidth
\epsfbox[76 234 488 544]{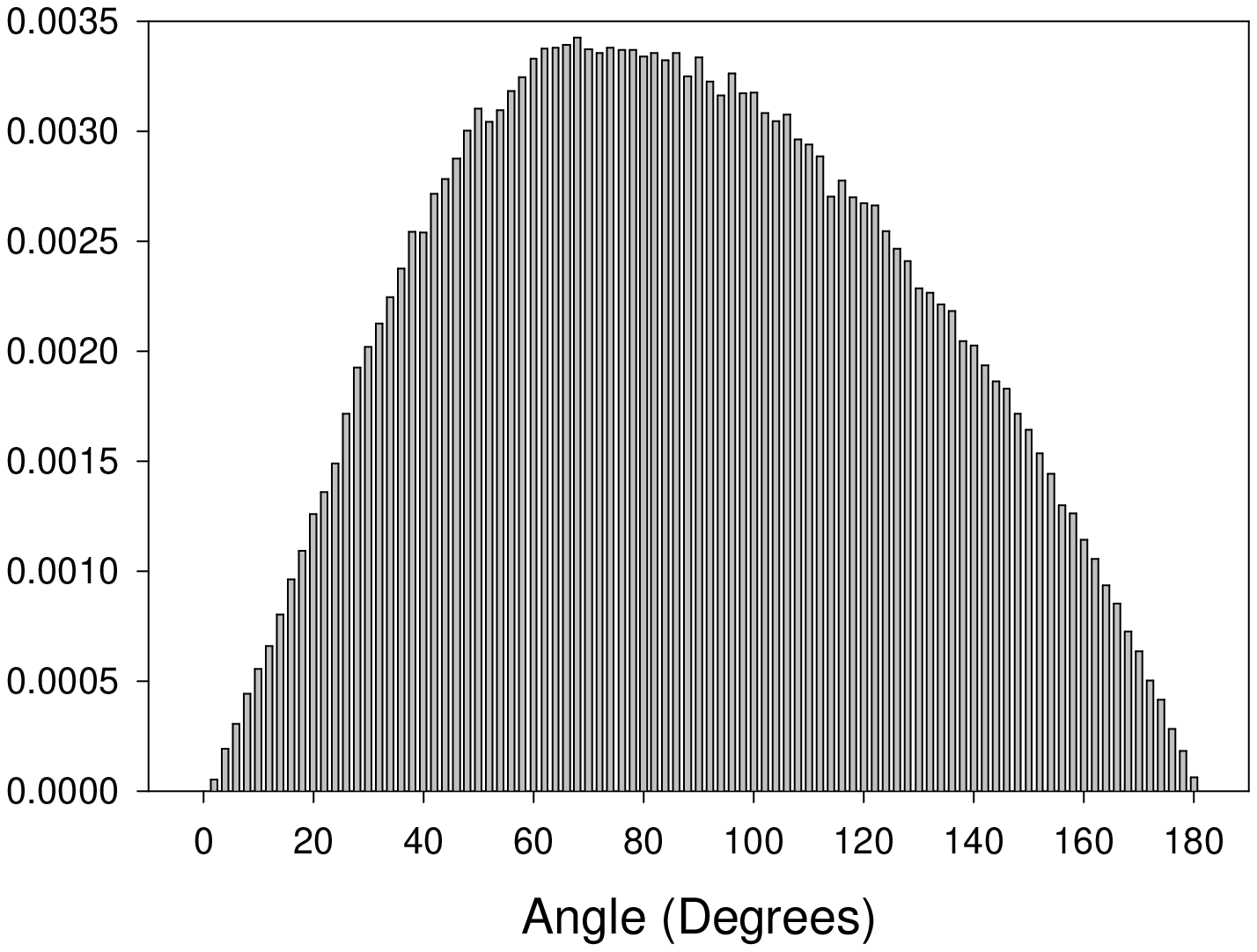}
\begin{minipage}{0.6\textwidth}
(a)
\end{minipage}
\end{figure}

\vspace{0.5 cm}

\begin{figure}
\centering
\epsfxsize=0.6\textwidth
\epsfbox[76 234 488 544]{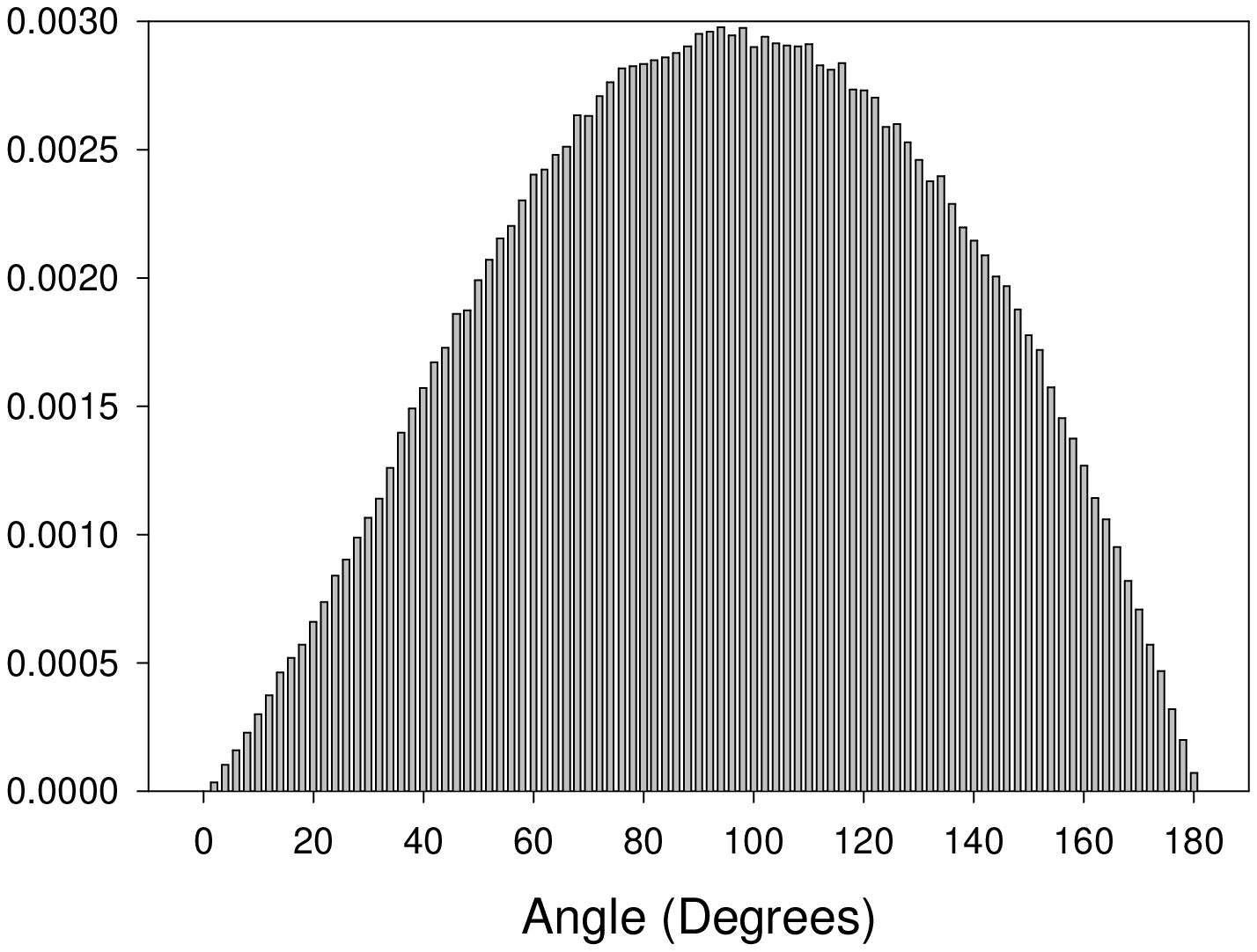}
\begin{minipage}{0.6\textwidth}
(a)
\end{minipage}
\end{figure}

\vspace{0.5 cm}

\end{document}